\documentclass[a4paper,twocolumn,11pt,accepted=2026-01-23]{quantumarticle}
\pdfoutput=1
\usepackage[utf8]{inputenc}
\usepackage[english]{babel}
\usepackage[T1]{fontenc}
\usepackage{amsmath}
\usepackage{amssymb}
\usepackage{pgfplots}
\pgfplotsset{compat=1.17}
\usepackage[numbers,sort&compress]{natbib}

\usepackage{tikz}
\usepackage{url}
\usepackage{graphicx}
\PassOptionsToPackage{compatibility=false}{caption}
\usepackage{subcaption}
\usepackage{placeins}
\usepackage{multirow}
\usepackage{verbatim}
\usepackage{booktabs}
\usepackage{lipsum} % For dummy text
\usepackage{enumitem}
\usepackage{adjustbox}
\usepackage{quantikz}
\usepackage{amsthm}

\usepackage{xcolor}
\usepackage{hyperref}
\DeclareCaptionLabelFormat{blue}{\textcolor{blue}{#1~#2}}

\makeatletter
\newtheoremstyle{blueDefinition}
  {3pt}   % Space above
  {3pt}   % Space below
  {}      % Body font
  {}      % Indent amount
  {\color{black}\bfseries} % Theorem head font (blue!)
  {.}     % Punctuation after head
  {.5em}  % Space after head
  {}      % Head spec
\makeatother

\theoremstyle{blueDefinition}
\newtheorem{bluedefinition}{Definition}

\newcommand{\mH}{\mathcal{H}}

\newcommand{\newtext}[1]{\textcolor{black}{#1}}

\begin{document}

\title{Quantum Architecture Search with Unsupervised Representation Learning}

\author{Yize Sun}
\thanks{Yize Sun and Zixin Wu contributed equally to this work.}
\affiliation{Ludwig-Maximilians-University Munich, Munich, 80539 Munich, Germany}
\affiliation{Siemens AG, 81739 Munich, Germany}
\affiliation{MCML, 80538 Munich, Germany}
\orcid{0009-0007-2921-2858}

\author{Zixin Wu}
\affiliation{Ludwig-Maximilians-University Munich, Munich, 80539 Munich, Germany}

\author{Volker Tresp}
\affiliation{Ludwig-Maximilians-University Munich, 
Munich, 80539 Munich, Germany}
\affiliation{Siemens AG, 81739 Munich, Germany}
\affiliation{MCML, 80538 Munich, Germany}

\author{Yunpu Ma}
\affiliation{Ludwig-Maximilians-University Munich, Munich, 80539 Munich, Germany}
\affiliation{MCML, 80538 Munich, Germany}
\email{cognitive.yunpu@gmail.com}

\maketitle
\begin{abstract}
Unsupervised representation learning presents new opportunities for advancing Quantum Architecture Search (QAS) on Noisy Intermediate-Scale Quantum (NISQ) devices. QAS is designed to optimize quantum circuits for Variational Quantum Algorithms (VQAs). Most QAS algorithms tightly couple the search space and search algorithm, typically requiring the evaluation of numerous quantum circuits, resulting in high computational costs and limiting scalability to larger quantum circuits. Predictor-based QAS algorithms mitigate this issue by estimating circuit performance based on structure or embedding. However, these methods often demand time-intensive labeling to optimize gate parameters across many circuits, which is crucial for training accurate predictors. Inspired by the classical neural architecture search algorithm \textit{Arch2vec}, we investigate the potential of unsupervised representation learning for QAS without relying on predictors. Our framework decouples unsupervised architecture representation learning from the search process, enabling the learned representations to be applied across various downstream tasks. Additionally, it integrates an improved quantum circuit graph encoding scheme, addressing the limitations of existing representations and enhancing search efficiency. This predictor-free approach removes the need for large labeled datasets. During the search, we employ REINFORCE and Bayesian Optimization to explore the latent representation space and compare their performance against baseline methods. We further validate our approach by executing the best-discovered MaxCut circuits on IBM's \texttt{ibm\_sherbrooke} quantum processor, confirming that the architectures retain optimal performance even under real hardware noise. Our results demonstrate that the framework efficiently identifies high-performing quantum circuits with fewer search iterations. 

Researchers and practitioners in quantum machine learning, quantum architecture search, and quantum circuit optimization, particularly those interested in applying unsupervised learning techniques to improve efficiency and scalability of circuit design for near-term quantum devices (NISQ).
\end{abstract}

\section{Introduction}
Quantum Computing has made significant progress over the past decades. Advances in quantum hardware and new quantum algorithms have demonstrated potential advantages~\citep{stein2023benchmarking} over classical computers in various tasks, such as image processing~\citep{wang2022review}, reinforcement learning~\citep{skolik2022quantum}, knowledge graph embedding~\citep{ma2019variational}, and network architecture search~\citep{zhang2022differentiable, giovagnoli2023qneat, du2022quantum}. However, the scale of quantum computers is still limited by environmental noise, which leads to unstable performance. These noisy intermediate-scale quantum (NISQ) devices lack fault tolerance, which is not expected to be achieved in the near future~\citep{preskill2018quantum}. The variational quantum algorithm (VQA), a hybrid quantum algorithm that utilizes quantum operations with adjustable parameters, is considered a leading strategy in the NISQ era~\citep{cerezo2021variational}. In VQA, the parameterized quantum circuit (PQC) with trainable parameters is viewed as a general paradigm of quantum neural networks and has achieved notable success in quantum machine learning. These parameters control quantum circuit operations, adjusting the distribution of circuit output states, and are updated by a classical optimizer based on a task-specific objective function. Although VQA faces challenges such as Barren Plateaus (BP) and scalability issues, it has demonstrated the potential to improve performance across various domains, including image processing, combinatorial optimization, chemistry, and physics~\citep{9996795, Amaro_2022, tilly2022variational}. One example of a VQA is the variational quantum eigensolver (VQE)~\citep{peruzzo2014variational, tilly2022variational}, which approximates the ground state and offers flexibility for quantum machine learning. We are considering using VQE to evaluate the performance of certain quantum circuits.

Unsupervised representation learning seeks to discover hidden patterns or structures within unlabeled data, a well-studied problem in computer vision research~\citep{radford2015unsupervised}. One common approach is the autoencoder, which is effective for feature representation. It consists of an encoder and decoder, which first maps images into a compact feature space and then decodes them to reconstruct similar images. Beyond images, autoencoders can also learn useful features from graphs, such as encoding and reconstructing directed acyclic graphs (DAGs) or neural network architectures~\citep{yan2020does, zhang2019d, pan2018adversarially, wang2016structural}. In most research, architecture search and representation learning are coupled, which results in inefficient searches heavily dependent on labeled architectures that require numerous evaluations. The \textit{Arch2vec} framework aims to decouple representation learning from architecture search, allowing downstream search algorithms to operate independently~\citep{yan2020does}. This decoupling leads to a smooth latent space that benefits various search algorithms without requiring extensive labeling. %This idea inspires us in this work, and we try to find if the decoupling can help quantum circuit architectures.

Quantum architecture search (QAS) or quantum circuit architecture search is a framework for designing quantum circuits efficiently and automatically, aiming to optimize circuit performance~\citep{du2022quantum}. Various algorithms have been proposed for QAS~\citep{zhang2022differentiable, du2022quantum, zhang2021neural, he2023gsqas, giovagnoli2023qneat}. However, most algorithms combine the search space and search algorithm, leading to inefficiency and high evaluation costs. The effectiveness of the search algorithm often depends on how well the search space is defined, embedded, and learned. Finding a suitable circuit typically requires evaluating different architectures many times. Although predictor-based QAS~\cite{he2023gsqas} can separate representation learning from the search algorithm, it often relies on labeling different architectures via evaluation, and the training performance depends heavily on the quantity and quality of evaluations and the embedding. In this work, \newtext{we decouple unsupervised architecture representation learning from the search process, enabling learned representations to be reused across different downstream tasks. This design is motivated by prior observations that the structure of parameterized quantum circuits strongly influence their expressibility, entanglement properties, and ultimately the performance of variational quantum algorithms~\cite{sim2019expressibility, Du_2020, benedetti2019parameterized}. Inspired by the success of Arch2vec, our approach learns structure-preserving representations of PQC architectures in an unsupervised manner, encouraging circuits with similar structural properties to occupy nearby regions in latent space, which empirically correlates with performance similarity even without labels.} We seek to explore whether decoupling can embed quantum circuit architectures into a smooth latent space, benefiting predictor-free QAS algorithms. We summarise our contributions as follows:

\begin{itemize}
\item We have successfully incorporated decoupling into unsupervised architecture representation learning within QAS, significantly improving search efficiency and scalability. By applying REINFORCE and Bayesian optimization directly to the latent representation, we eliminate the need for a predictor trained on large labeled datasets, thereby reducing prediction uncertainty.%(item 2. improved encoding)
\item Our proposed quantum circuit encoding scheme overcomes limitations in existing representations, enhancing search performance by providing more accurate and effective embeddings.
\item Extensive experiments on quantum machine learning tasks, including quantum state preparation, max-cut, and quantum chemistry \citep{liang2019quantum, poljak1995solving, tilly2022variational}, confirm the effectiveness of our framework on simulator and real quantum hardware. The pre-trained quantum architecture embeddings significantly enhance QAS across these applications.
\end{itemize}

\section{Related Work}
\paragraph{Unsupervised Graph Representation Learning.} Graph data is becoming a crucial tool for understanding complex interactions between real-world entities, such as biochemical molecules \citep{jiang2021learning}, social networks \citep{shen2023uniskgrep}, purchase networks from e-commerce platforms \citep{li2021towards}, and academic collaboration networks \citep{newman2001structure}. Graphs are typically represented as discrete data structures, making it challenging to solve downstream tasks due to large search spaces.
Our work focuses on unsupervised graph representation learning, which seeks to embed graphs into low-dimensional, compact, and continuous representations without supervision while preserving the topological structure and node attributes. In this domain, approaches such as those proposed by \cite{perozzi2014deepwalk, wang2016structural, grover2016node2vec, tang2015line} use local random walk statistics or matrix factorization-based objectives to learn graph representations. Alternatively, methods like \cite{kipf2016variational, hamilton2017inductive} reconstruct the graph's adjacency matrix by predicting edge existence, while others, such as \cite{velivckovic2018deep, sun2019infograph, peng2020graph}, maximize the mutual information between local node representations and pooled graph representations.
Additionally, \cite{xu2019powerful} investigate the expressiveness of Graph Neural Networks (GNNs) in distinguishing between different graphs and introduce Graph Isomorphism Networks (GINs), which are shown to be as powerful as the Weisfeiler-Lehman test \citep{leman1968reduction} for graph isomorphism. Inspired by the success of \textit{Arch2vec} \citep{yan2020does}, which employs unsupervised graph representation learning for classical neural architecture search (NAS), we adopt GINs to encode quantum architecture structures, as quantum circuit architectures can also be represented as DAGs.

\paragraph{Quantum Architecture Search (QAS).} As discussed in the previous section, PQCs are essential as ansatz for various VQAs \citep{benedetti2019parameterized}. The expressive power and entangling capacity of PQCs play a crucial role in their optimization performance \citep{sim2019expressibility}. Poorly designed ansatz can suffer from limited expressive power or entangling capacity, making it difficult to reach the global minimum for an optimization problem. Moreover, such ansatz may be more prone to noise \citep{stilck2021limitations}, inefficiently utilize quantum resources, or lead to barren plateaus that hinder the optimization process \citep{mcclean2018barren, wang2021noise}. To address these challenges, QAS has been proposed as a systematic approach to identify optimal PQCs.
\newtext{Recent progress in QAS can be broadly grouped into several methodological families. 
\emph{Reinforcement-learning–based approaches}, such as QCDS, RL-QMLAS, RL-VQC, RL-QAOA \citep{Pirhooshyaran2021, kuo2021quantum, ostaszewski2021reinforcement, Khairy_2020}, and DRL-based circuit rewriting \citep{Foesel2021QuantumCircuitOptimization}, use policy agents to explore discrete circuit spaces or apply rewrite rules to improve existing circuits. 
\emph{Differentiable or gradient-based QAS methods} have been proposed by relaxing discrete gate choices into continuous mixtures~~\cite{zhang2022differentiable, pmlr-v202-wu23v}. While effective in settings where circuit architectures themselves can be continuously parameterized and optimized via gradients, these approaches rely on relaxation-specific assumptions and gradient access at the architecture level. In contrast, our work treats PQC architectures as discrete objects that are evaluated via task-level VQA executions, without assuming differentiability with respect to architectural choices. Exploration is instead carried out through a continuous latent embedding learned in an unsupervised manner. This distinction makes differentiable QAS complementary rather than directly comparable to the search setting considered in this work.\emph{Evolutionary methods} such as QAS-EA, EQAS, EQAS-PQC, QNEAT, MoG-VQE, and NACL \citep{du2022quantum, cincio2021machine, chivilikhin2020mogvqemultiobjectivegeneticvariational, zhang2022evolutionary, ding2022evolutionary, giovagnoli2023qneat, huang2022robust, franken2022quantum} evolve circuit structures via mutation--selection mechanisms, often incorporating multiobjective criteria or hardware-noise awareness. 
Other approaches include \emph{greedy search} \citep{tang2021qubit, PhysRevResearch.6.033033} and \emph{Bayesian optimization} \citep{duong2022quantum}}. 
However, these methods require the evaluation of numerous quantum circuits, which is both time-consuming and computationally expensive. To mitigate this issue, predictor-based approaches \citep{zhang2021neural, he2023gnn} have been introduced, but they also face limitations. These approaches rely on large sets of labeled circuits to train predictors with generalized capabilities and introduce additional uncertainty into the search process, necessitating the reevaluation of candidate circuits. 

\newtext{While these approaches explore circuit architectures directly in discrete or relaxed spaces, they do not leverage unsupervised representation learning to decouple structure learning from performance-driven search, which is the focus of our work. In contrast, our method differs from these lines of work in that the search is not performed directly in the discrete circuit space, nor via differentiable relaxation, but instead through a latent embedding learned in an unsupervised manner. This representation greatly reduces the combinatorial search complexity while eliminating the need for labeling or predictor training, allowing Reinforcement Learning (RL) and Bayesian Optimization (BO) to operate efficiently in a continuous latent space.}
\section{QAS with Unsupervised Representation Learning}

\begin{figure*}[ht]
\begin{center}
    \centering
    \begin{subfigure}{0.45\linewidth}
    \includegraphics[width = \linewidth]{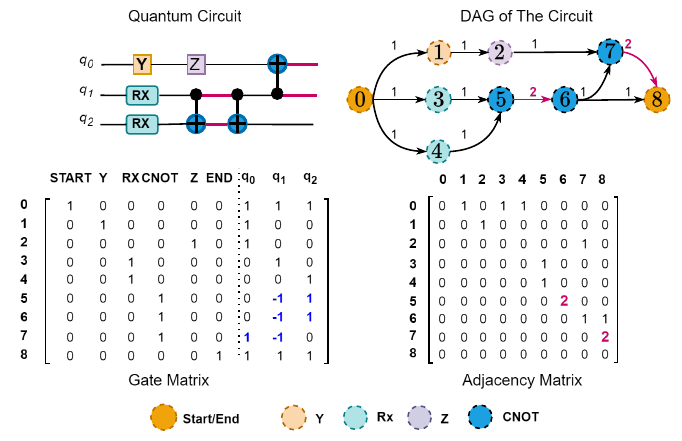} %
    \caption{Architecture encoding scheme}
    \label{algorithm-1}
    \end{subfigure}
    \hfill
    \begin{subfigure}{0.5\linewidth}
    \includegraphics[width = \linewidth]{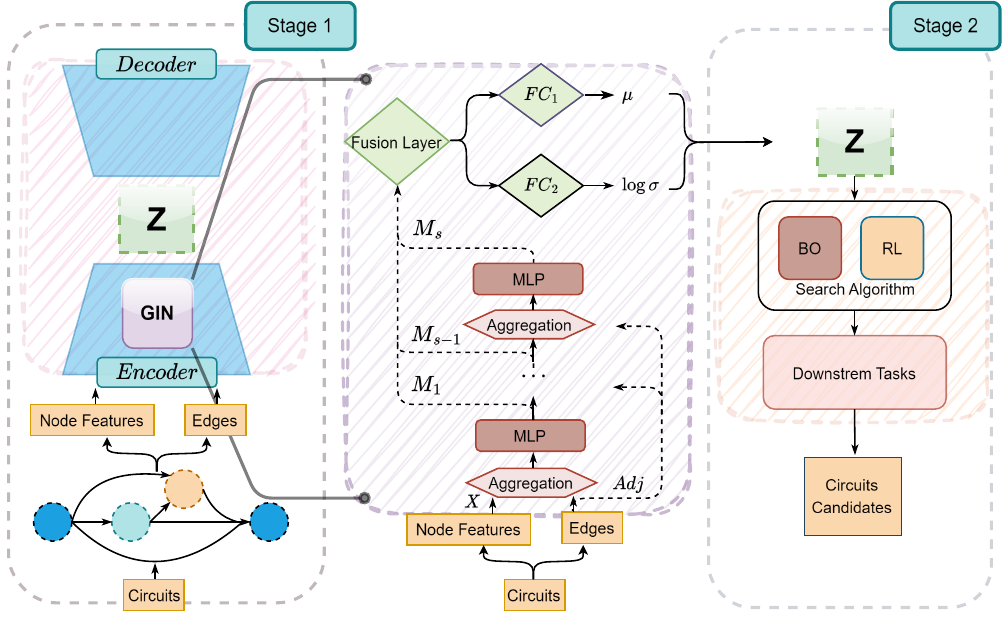} %
    \caption{Representation learning and search process}
    \label{algorithm-2}
    \end{subfigure}

    \caption{Illustration of our algorithm. In Figure \ref{algorithm-1}, each circuit's architecture is first transformed into a DAG and represented by two matrices. Each row of the gate matrix corresponds to a node in the graph, with one-hot encoding used to indicate the node type, and additional columns encoding position information, such as the qubits the gate acts on. For two-qubit gates, $-1$ and $1$ represent the control and target qubits, respectively. The weights in the adjacency matrix reflect the number of qubits involved in each interaction. In Figure \ref{algorithm-2}, the left side depicts the process of representation learning, where $Z$ represents the latent space of circuit architectures. In the middle, the encoder is shown as the mechanism used to learn this latent space. On the right, Bayesian optimization (BO) and reinforcement learning (RL) are employed to explore the latent space for various quantum machine learning tasks. The algorithm ultimately outputs a set of candidate circuits.}
    \label{algorithm} 
\end{center}
\end{figure*}
In this work, we present our method, as illustrated in Figure \ref{algorithm}, \newtext{which consists of two independent components: (1) unsupervised representation learning and (2) predictor-free architecture search.} The search space is defined by the number of gates in a circuit and an operation pool comprising general gate types such as ${\texttt{X, Y, Z, H, Rx, Ry, Rz, U3, CNOT, CY, CZ}}$. A random generator creates a set of circuit architectures based on predefined parameters, including the number of qubits, the number of gates, and the maximum circuit depth. These architectures are then encoded into two matrices and input into the autoencoder. The autoencoder independently learns a latent distribution from the search space and produces pre-trained architecture embeddings for the search algorithms. The evaluation strategy takes the circuit architectures generated by the search algorithm and returns a performance assessment. For evaluating circuit architectures, we use the ground state of a Hamiltonian for max-cut and quantum chemistry problems, and fidelity for quantum state preparation tasks.

\begin{bluedefinition} We are given a circuit created by $m$ gate types, $h$ gates and $g$ qubits. Then, the circuit can be described by a DAG $G=\{V, E\}$ with $n=h+2=|V|$ gate nodes including START and END. The adjacency matrix of graph $G$ is summarized in $n\times n$ matrix $A$ and its gate matrix $X$ is in size of $n \times (m+2+g)$. We further introduce $d$-dimensional latent variables $z_i$ composing latent matrix $Z=\{z_1,..,z_K\}^T$.
\label{definition-1}
\end{bluedefinition}

\subsection{Circuit Encoding Scheme}
\newtext{Following Definition~\ref{definition-1}, we represent quantum circuits as DAGs, extending the encoding strategy used in $\mathcal{E}^{GSQAS}$ [49,20]. Each quantum gate in the circuit is treated as a node, and directed edges encode execution dependencies between gates acting on the same qubit. Because a quantum circuit is a sequence of quantum operations, we additionally introduce a global start node and a global end node so that all qubit lines originate from the former and terminate at the latter. This guarantees a consistent topological ordering and ensures that the resulting structure forms a DAG. From this DAG, we construct two complementary representations. The adjacency matrix $A$ captures the structural topology of the circuit, where $A_{ij}=1$ if there exists a directed edge from node $i$ to node $j$, and $A_{ij}=0$ otherwise. In parallel, we construct the gate matrix $X \in \mathbb{R}^{n \times (m+2+g)}$, where each row corresponds to a gate and is encoded using a one-hot or embedding-based representation of its gate type.}

\newtext{\paragraph{Limitations of prior encodings.} However, the original $\mathcal{E}^{GSQAS}$ encoding treats all occupied qubits uniformly and does not distinguish control and target roles in two-qubit operations (e.g., \texttt{CNOT}, \texttt{CZ}). This ambiguity may lead to representation loss and suboptimal circuit reconstruction. Moreover, the adjacency matrix in the original scheme assigns uniform edge weights, which fails to reflect the structural importance of multi-qubit interactions.}

\newtext{\paragraph{Enhanced encoding scheme.} To address these limitations, we introduce an enhanced encoding scheme. For two-qubit gates as shown in Figure~\ref{algorithm-1}, we explicitly encode operational roles by assigning $-1$ to the control qubit and $+1$ to the target qubit in the corresponding row of $X$. Additionally, we assign edge weights proportional to the number of qubits involved in the operation (e.g., weight $2$ for two-qubit gates) in the adjacency matrix. These modifications retain operational structure, qubit roles, and gate-type distinctions, enabling a more expressive and faithful circuit representation. As demonstrated in later sections, this refined encoding improves unsupervised representation learning and enhances the performance of latent-space architecture search.}

\subsection{\textcolor{black}{Unsupervised representation learning}}
\newtext{The objective of this stage is to learn latent representations of quantum
circuits without relying on performance labels or VQA evaluations. We jointly
train a GIN-based encoder and a fully connected decoder by maximizing the ELBO over 100{,}000 randomly generated circuits using only structural inputs $(A, X)$.}

\subsubsection{Preliminaries}
\newtext{Graph autoencoders (GAEs) provide a general framework for unsupervised graph
representation learning, consisting of an encoder that maps a graph to a latent
space and a decoder that reconstructs the graph from this representation.
A representative example is the variational graph autoencoder (VGAE), which
uses a graph convolutional network as the encoder and optimizes a variational
objective~\citep{kipf2016variational}.}

\newtext{In this work, we adopt the graph autoencoder paradigm but depart from the
standard VGAE design. Specifically, we employ a Graph Isomorphism Network
(GIN)~\citep{xu2019powerful} as the encoder, which has been shown to be
theoretically more expressive than common GCN-based encoders in distinguishing
graph structures. This choice is particularly important for quantum circuits,
where subtle differences in connectivity and gate ordering can lead to
significant differences in circuit behavior.}

\subsubsection{Encoder}

The encoder GIN maps the circuit structure and node features to a set of latent
variables $Z$. We approximate the posterior distribution over latent variables
as
\begin{align}
    q(Z \mid X, A) = \prod_{i=1}^{K} q(z_i \mid X, A),
\end{align}
where $q(z_i|X,A)=\mathcal{N}(z_i|\mu_i, \text{diag}(\sigma^2_i))$\text{.} \newtext{During representation learning, this variational posterior provides circuit-dependent latent encodings, while after training latent vectors are sampled from the prior $p(z)$ for architecture search.}

The GIN consists of $L$ message-passing layers, which generate intermediate node embeddings $M^{(s)}$ according to:
\begin{align}
    M^{(s)} &= \mathrm{MLP}^{(s)}\!\left((1+\epsilon^{(s)}) M^{(s-1)}
    + \hat{A} M^{(s-1)}\right), \nonumber \\
    s &= 1,2,\dots,L ,
    \label{eq:gin_update}
\end{align}
where $M^{(0)} = X$ denotes the initial node feature matrix derived from the gate encoding. \newtext{The term $\hat{A} M^{(s-1)}$ performs neighborhood aggregation by summing feature vectors of gates that are connected in the circuit graph, thereby propagating structural information along circuit dependencies. The term $(1+\epsilon^{(s)}) M^{(s-1)}$ introduces a learnable self-loop that controls the relative contribution of a node’s own features at layer $s$. The symmetrized adjacency matrix $\hat{A} = A + A^\top$ enables bidirectional message passing between connected gates. Finally, $\mathrm{MLP}^{(s)}$ applies a shared nonlinear transformation to the aggregated features to produce the updated node embeddings at layer $s$.}

After $L$ GIN layers, we aggregate the learned node embeddings to parameterize
the variational posterior. Instead of relying solely on the final GIN layer, we
introduce a fusion layer that aggregates information from all layers to improve
representation robustness. The mean and standard deviation are computed as
$\mu = \mathrm{GIN}_{\mu}(X, \hat{A}) = \mathrm{FC}_1(M^{(L)})$ and
$\sigma = \mathrm{FC}_2(M^{(L)})$, respectively. Latent variables are then
sampled using the reparameterization trick
$z_i = \mu_i + \sigma_i \cdot \epsilon_i$, where $\epsilon_i \sim \mathcal{N}(0, I)$.

For all experiments, we use $L = 5$ GIN layers, a latent dimension of $16$, and a
hidden dimension of $128$ for the GIN encoder. Additional hyperparameter details
are provided in Appendix~\ref{pretraining_parameters}.

\subsubsection{Decoder}
\newtext{The decoder takes the sampled latent variables $Z$ as input to reconstruct both the adjacency matrix $A$ and the gate matrix $X = [X^t, X^q]$, where $X^t$ corresponds to the one-hot encoding of the gate type for each node, and $X^q$ encodes the qubit positions associated with each gate.} The generative process is summarized as follows:
\begin{align}
     p(A|Z)&=\prod^{K}_{i=1}\prod^{K}_{j=1}p(A_{ij}|z_i,z_j),\\
     &\text{ with } p(A_{ij}|z_i,z_j)=\text{ReLU}_j(F_{1}(z^T_iz_j))\text{,}\\
     p(X|Z)&=\prod^K_{i=1}p(x_i|z_i)\text{,}\\
     &\text{ with } p(x^t_i|z_i)=\text{softmax}(F_{2}(z_i))\text{, }\\
     &p(x^q_i|z_i)=\text{tanh}(F_{2}(z_i))\text{,}
\end{align}
where both $F_{1}$ and $F_{2}$ are trainable linear functions.

\subsubsection{Objective Function}
The weights in the encoder and decoder are optimized by maximizing the evidence lower bound (ELBO) $\mathcal{L}$, which is defined as:
\begin{align}
    \mathcal{L}=E_{q(Z|X,A)}[\log p(X^{\text{type}},X^{\text{qubit}},A|Z)]\nonumber\\
    -\text{KL}[(q(Z|X,A))||p(Z)]\text{,}
\end{align}
where KL$[q(\cdot)||p(\cdot)]$ represents the Kullback-Leibler (KL) divergence between $q(\cdot)$ and $p(\cdot)$. We further adopt a Gaussian prior $p(Z)=\prod_i \mathcal{N}(z_i|0,I)$. The weights are optimized using minibatch gradient descent, with a batch size of 32.

\newtext{The representation learning stage is fully unsupervised and operates solely on 
circuit structure $(A, X)$. Because the GIN encoder aggregates local gate 
patterns and connectivity, circuits with similar structural characteristics 
tend to yield similar node embeddings, and the reconstruction objective 
encourages these embeddings to form a smooth latent space. This intuition 
aligns with prior studies showing that circuit structure strongly influences 
expressibility and entangling capability, and thereby the expressive behavior 
of PQCs in variational algorithms~\cite{sim2019expressibility, Du_2020, benedetti2019parameterized}. 
As observed in Arch2vec, a variational graph isomorphism autoencoder trained 
only on architectural structure (without using performance labels) encourages 
architectures with similar connections to cluster in latent space, and 
architectures with similar performance tend to occupy nearby regions in that 
space~\cite{yan2020does}. 
As a result, shared structural patterns often correlate with performance-relevant 
properties in variational algorithms, an effect that is also reflected in the 
smooth organization of circuits observed in our latent space visualizations (Figure~2).}

\subsection{\textcolor{black}{Predictor-free architecture search}}
\newtext{In the search stage, we freeze the pretrained encoder and explore the latent space using RL/BO. Candidate architectures are evaluated only for reward computation, resulting in a predictor-free search process.}

\subsubsection{Reinforcement Learning (RL)}
%TODO check with zixin REINFORCE, latent vector, check layer LSTM
After conducting initial trials with PPO~\citep{schulman2017proximal} and A2C~\citep{huang2022a2c}, we adopt REINFORCE~\citep{Williams:92} as a more effective reinforcement learning algorithm for architecture search. In this approach, the environment's state space consists of pre-trained embeddings, and the agent uses a one-cell LSTM as its policy network. The agent selects an action, corresponding to a sampled latent vector based on the distribution of the current state, and transitions to the next state based on the chosen action. \newtext{For the MaxCut and quantum chemistry tasks, we adopt an energy convention in which the ground-state energy $E_{\text{ground}}$ is negative and $0$ corresponds to a trivial upper bound. The reward used in the REINFORCE search is defined as}

\begin{equation}
\color{black}
r = \frac{E_{\text{candidate}}}{E_{\text{ground}}}.
\end{equation}

\newtext{Under this convention, $r \in [0,1]$ for valid evaluations, where $r=1$
indicates reaching the ground state and $r=0$ corresponds to
$E_{\text{candidate}} = 0$. During early optimization or in the presence of
numerical noise, $E_{\text{candidate}}$ may become slightly positive, producing a negative ratio; in such cases, we clip the reward to $0$ to ensure a stable
and interpretable reward signal.} For the state preparation task, circuit fidelity is used as the reward.
We employ an adaptive batch size, with the number of steps per training epoch determined by the average reward of the previous epoch. Additionally, we use a linear adaptive baseline, defined by the formula $B = \alpha \cdot B + (1 - \alpha) \cdot R_{avg}$, where $B$ denotes the baseline, $\alpha$ is a predefined value in the range [0,1], and $R_{avg}$ is the average reward. Each run in this work involves 1000 searches.

\subsubsection{Bayesian Optimization (BO)}
As another search strategy used in this work without labeling, we employ Deep Networks for Global Optimization (DNGO)\citep{snoek2015scalable} in the context of BO. We adopt a one-layer adaptive BO regression model with a basis function extracted from a feed-forward neural network, consisting of 128 units in the hidden layer, to model distributions over functions. Expected Improvement (EI)\citep{conf/ifip/Mockus77} is selected as the acquisition function. EI identifies the top-k embeddings for each training epoch, with a default objective value of 0.9. The training begins with an initial set of 16 samples, and in each subsequent epoch, the top-k architectures proposed by EI are added to the batch. The network is retrained for 100 epochs using the architectures from the updated batch. This process is iterated until the predefined number of search iterations is reached.

\section{Experimental Results}
To demonstrate the effectiveness and generalization capability of our approach, we conduct experiments on three well-known quantum computing applications: quantum state preparation, max-cut, and quantum chemistry. For each application, we start with a simple example involving 4 qubits and then progress to a more complex example with 8 qubits. We utilize a random generator to create 100,000 circuits as the search space, and all experiments are performed on a noise-free simulator during the search process. Detailed settings are provided in Appendix \ref{application_setting}. We begin by evaluating the model's pre-training performance for unsupervised representation learning (Appendix~\ref{pretraning_performance}), followed by an assessment of QAS performance based on the pre-trained latent representations (Appendix~\ref{qas_performance}).

\subsection{Pre-training Performance}
\label{pretraning_performance}
\textbf{Observation (1):} GAE and VGAE \citep{kipf2016variational} are two popular baselines for NAS. In an attempt to find models capable of capturing superior latent representations of quantum circuit architectures, we initially applied these two well-known models. However, due to the increased complexity of quantum circuit architectures compared to neural network architectures, these models failed to deliver the expected results. In contrast, models based on GINs \citep{xu2019powerful} successfully obtained valid latent representations, attributed to their more effective neighbor aggregation scheme.
Table \ref{model_performance} presents a performance comparison between the original model using the $\mathcal{E}^{GSQAS}$ encoding and the improved model with our enhanced encoding for 4, 8, and 12 qubit circuits, evaluated across five metrics: Accuracy${ops}$, which measures the reconstruction accuracy of gate types in the gate matrix for the held-out test set; Accuracy${qubit}$, which reflects the reconstruction accuracy of qubits that the gates act on; Accuracy${adj}$, which measures the reconstruction accuracy of the adjacency matrix; Falpos${mean}$, which represents the mean false positives in the reconstructed adjacency matrix; and KLD (KL divergence), which indicates the continuity and smoothness of the latent representation.
The results in the table indicate that the improved model with our enhanced encoding achieves comparable or better than the original. This improvement can be attributed to two factors: first, the new encoding better captures the specific characteristics of the circuits, and second, the fusion of outputs from multiple layers of GIN helps retain shallow information, resulting in more stable training.
\begin{table*} [ht]
    \footnotesize
    \centering
    \begin{tabular}{ l l|l l l l l}
    \hline
    \multirow{2}{*}{Qubit}&
    \multirow{2}{*}{Model}&
    \multicolumn{5}{c}{Metric}\\
    \cline{3-7}
    & & Accuracy$_{ops}$ & Accuracy$_{qubit}$ & Accuracy$_{adj}$ & Falpos$_{mean}$ & KLD\\
    \hline
    4 & GSQAS & 99.99 & 99.99 & 99.91 & 100.00 & 0.061\\
    \hline
    4 & Ours & \textbf{100} & 99.97 & 98.89 & \textbf{23.41} & \textbf{0.045}\\
    \hline
    8 & GSQAS & 86.69 & 99.98 & 99.82 & 100.00 & 0.038\\
    \hline
    8 & Ours & \textbf{100} & 98.65 & 97.34 & \textbf{7.35} & \textbf{0.029}\\
    \hline
    12 & GSQAS & 86.69 & 99.94 & 99.70 & 100.00 & 0.028\\
    \hline
    12 & Ours & \textbf{98.67} & 99.14 & 97.79 & \textbf{4.75} & \textbf{0.022}\\
    \hline
    \end{tabular}
    \caption{Pretraining model performance of 4-, 8-, and 12-qubit circuits across the four metrics.}
    \label{model_performance}
\end{table*}

\begin{figure*}[ht]
    \centering
    % First row
    
    \begin{subfigure}{0.23\textwidth}
        \centering
        \includegraphics[width=\textwidth]{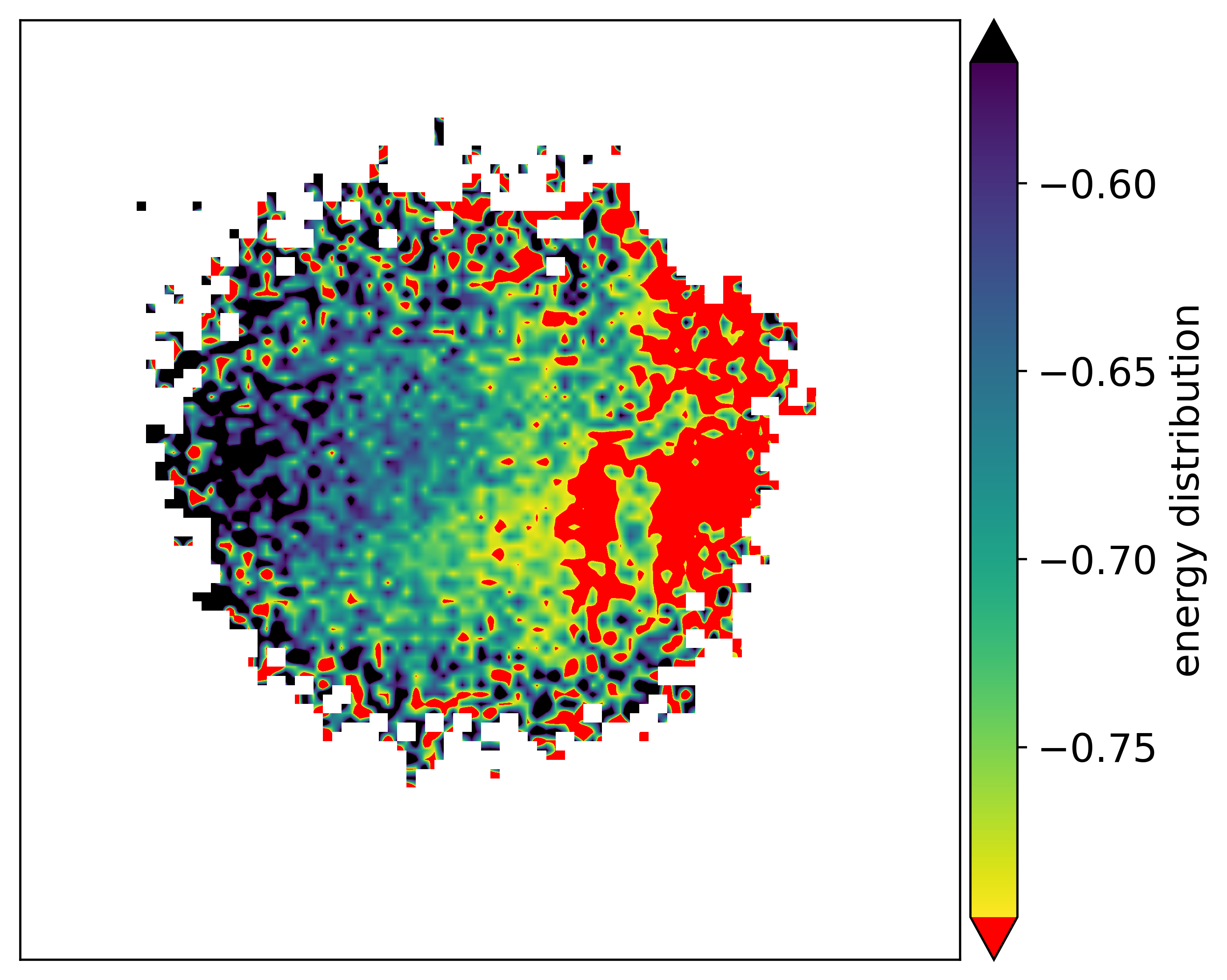}
        \caption{PCA$_4$ QC$_{H_2}$}
        \label{fig:PCA-4}
    \end{subfigure}
    %\hspace{0.02\textwidth}
    \begin{subfigure}{0.23\textwidth}
        \centering
        \includegraphics[width=\textwidth]{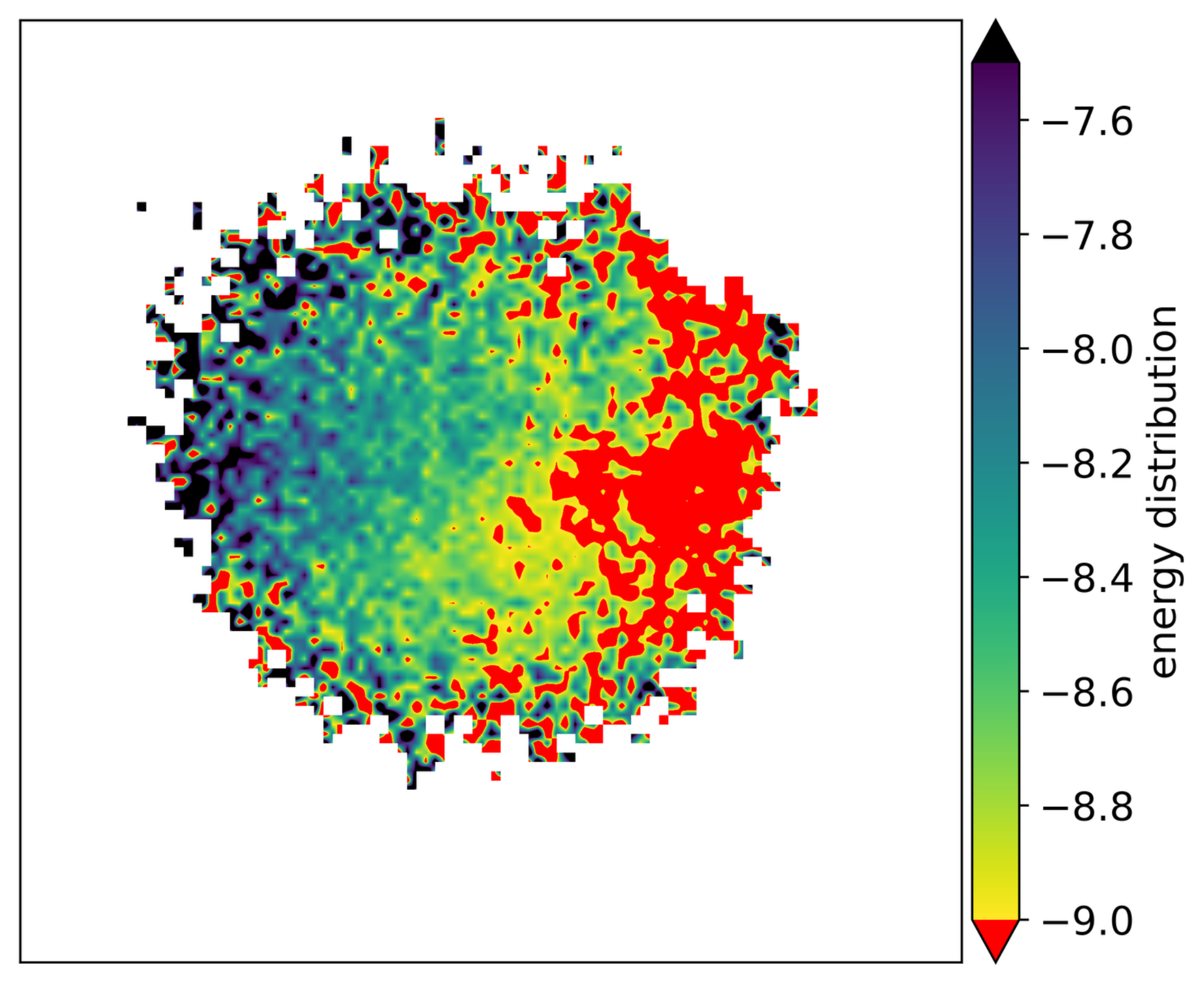}
        \caption{PCA$_4$ Max-cut}
        \label{fig:PCA-4-Max-cut}
    \end{subfigure}
    \begin{subfigure}{0.23\textwidth}
        \centering
        \includegraphics[width=\textwidth]{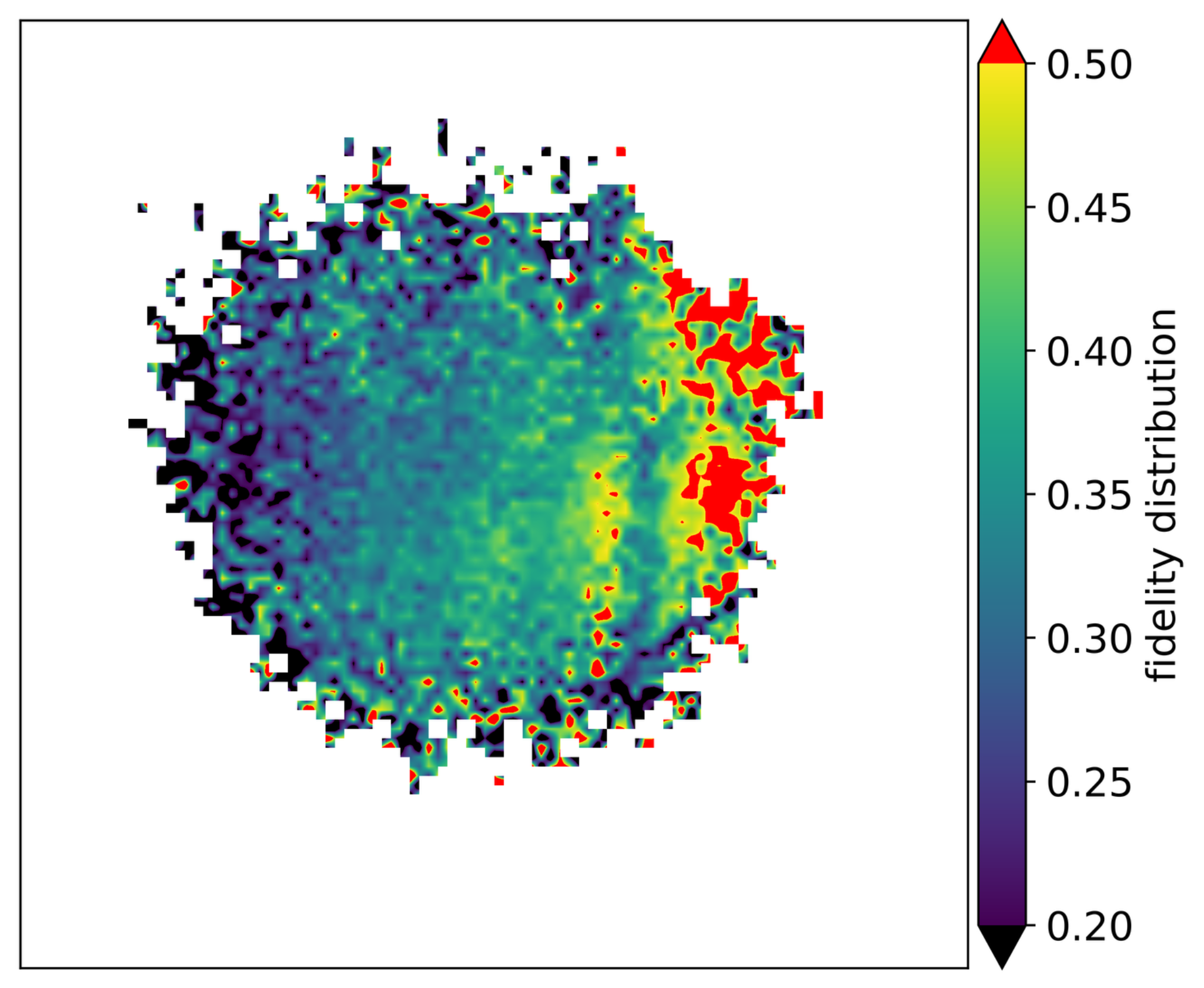}
        \caption{PCA$_4$ Fidelity}
        \label{fig:PCA-4-Fidelity}
    \end{subfigure}
    \begin{subfigure}{0.23\textwidth}
        \centering
        \includegraphics[width=\textwidth]{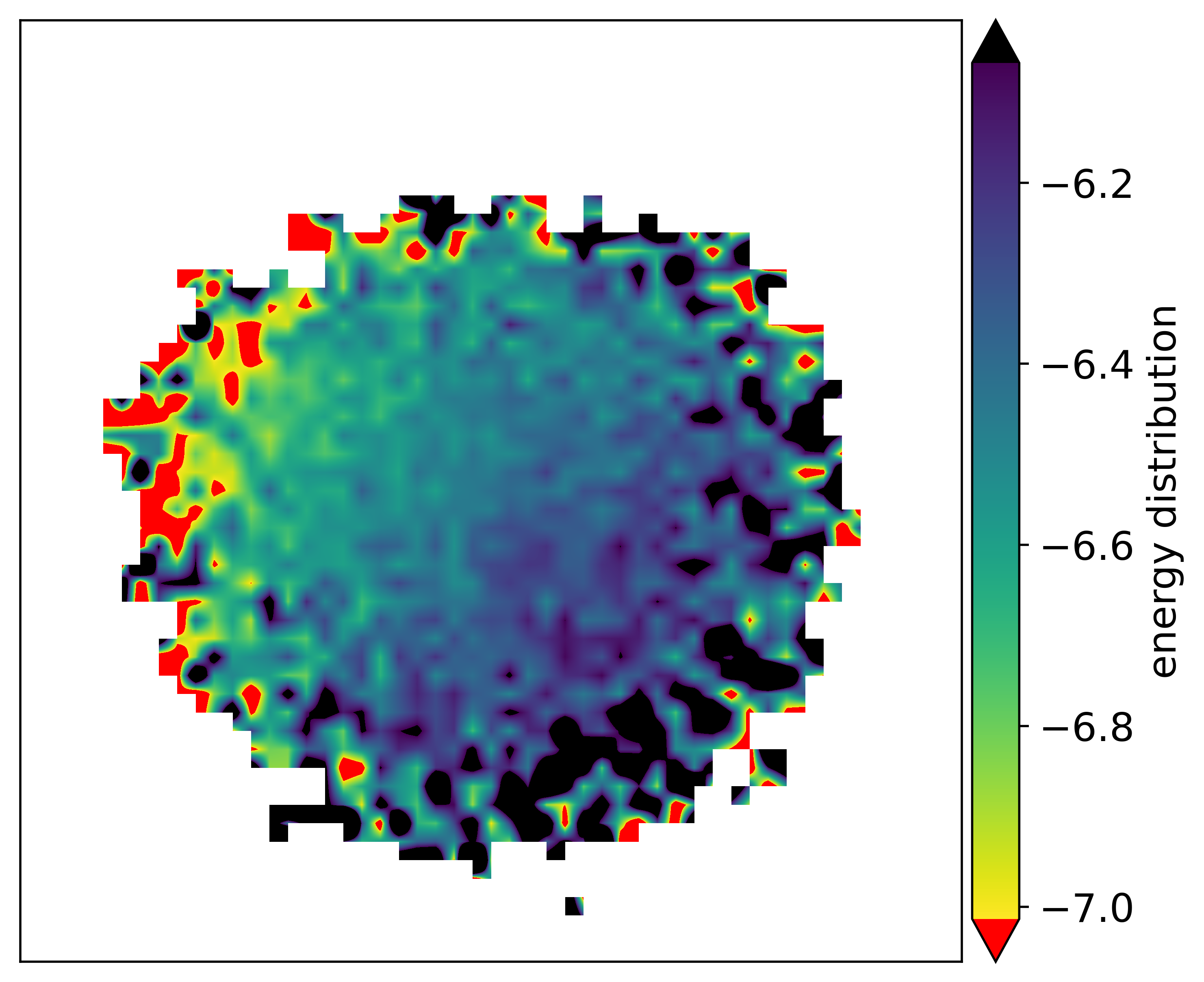}
        \caption{PCA$_{12}$ QC$_{LiH}$}
        \label{fig:PCA-12}
    \end{subfigure}
    
    % Second row
    \begin{subfigure}{0.23\textwidth}
        \centering
        \includegraphics[width=\textwidth]{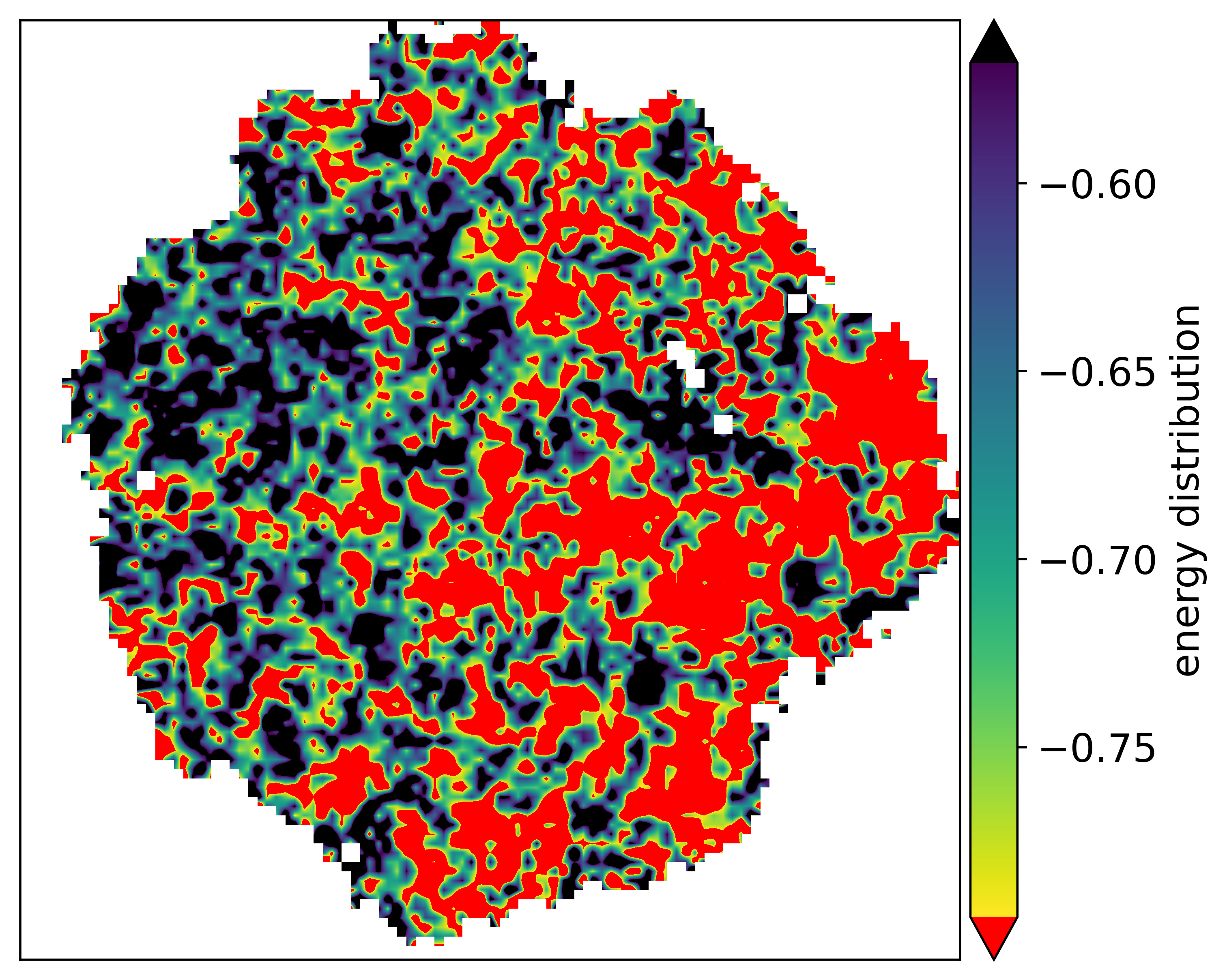}
        \caption{t-SNE$_4$ QC$_{H_2}$}
        \label{fig:t-SNE-4}
    \end{subfigure}
    %\hspace{0.02\textwidth}
    \begin{subfigure}{0.23\textwidth}
        \centering
        \includegraphics[width=\textwidth]{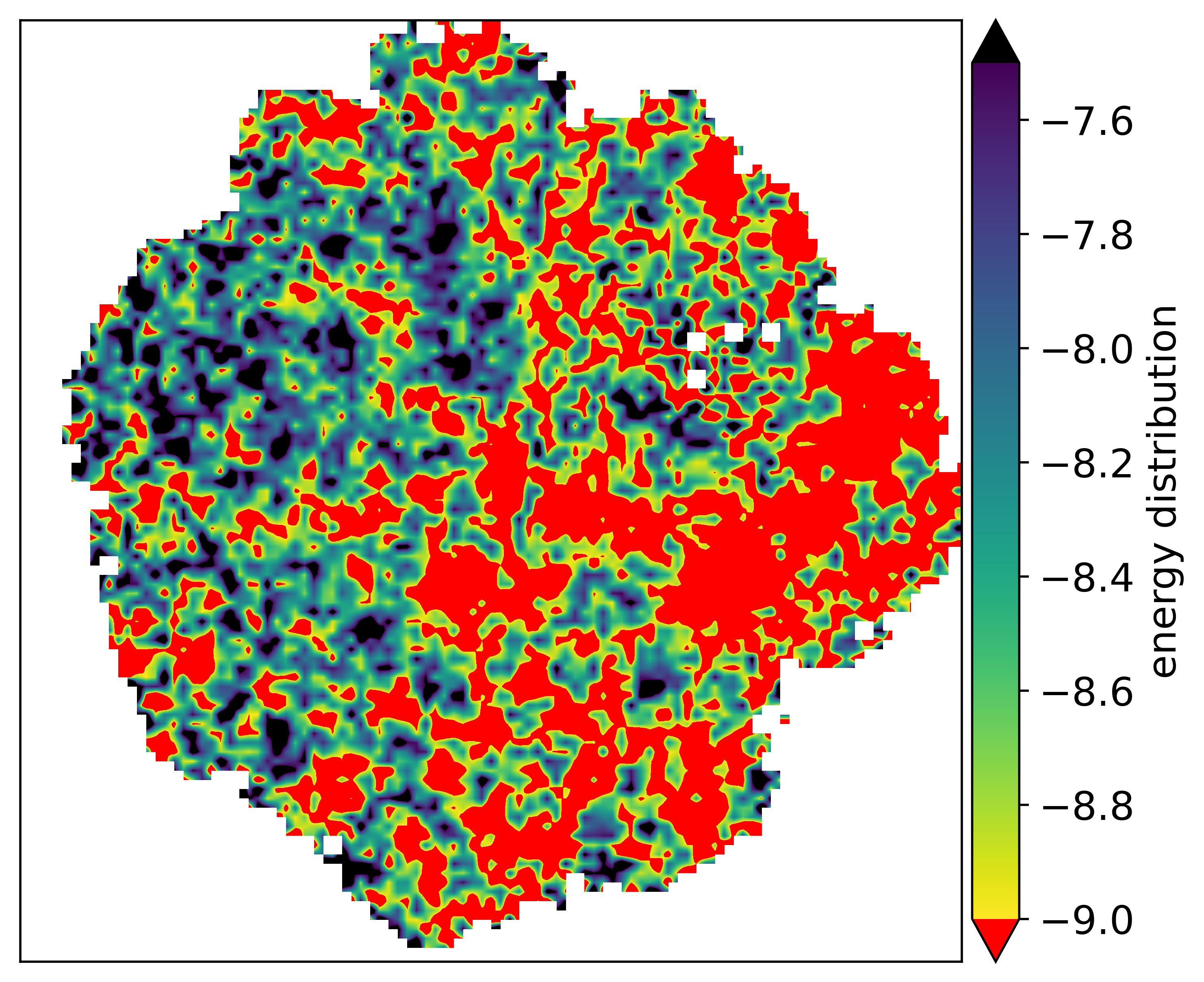}
        \caption{t-SNE$_4$ Max-cut}
        \label{fig:t-SNE4 Maxcut}
    \end{subfigure}
    \begin{subfigure}{0.23\textwidth}
        \centering
        \includegraphics[width=\textwidth]{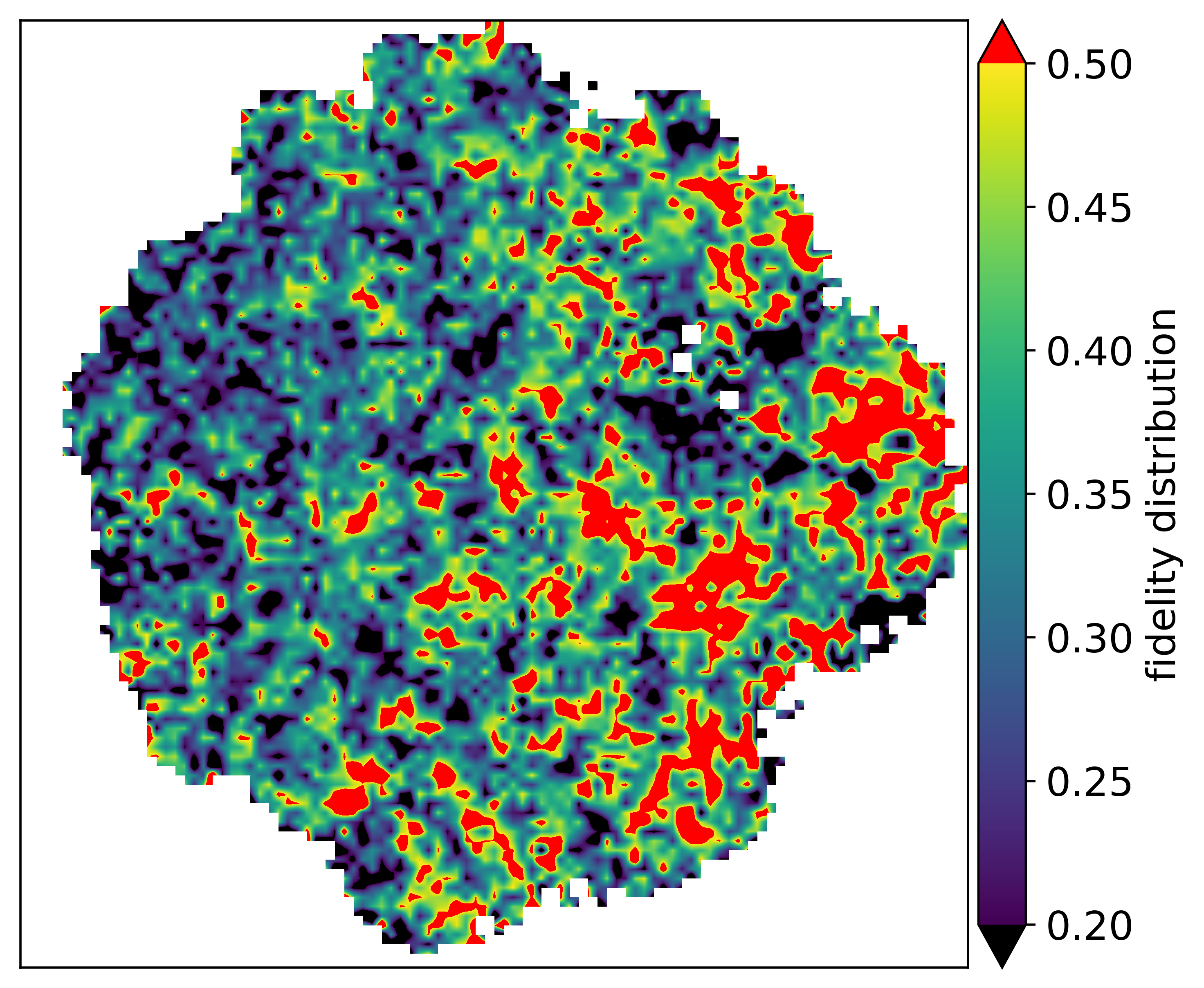}
        \caption{t-SNE$_4$ Fidelity}
        \label{fig:t-SNE4 Prev}
    \end{subfigure}
    \begin{subfigure}{0.23\textwidth}
        \centering
        \includegraphics[width=\textwidth]{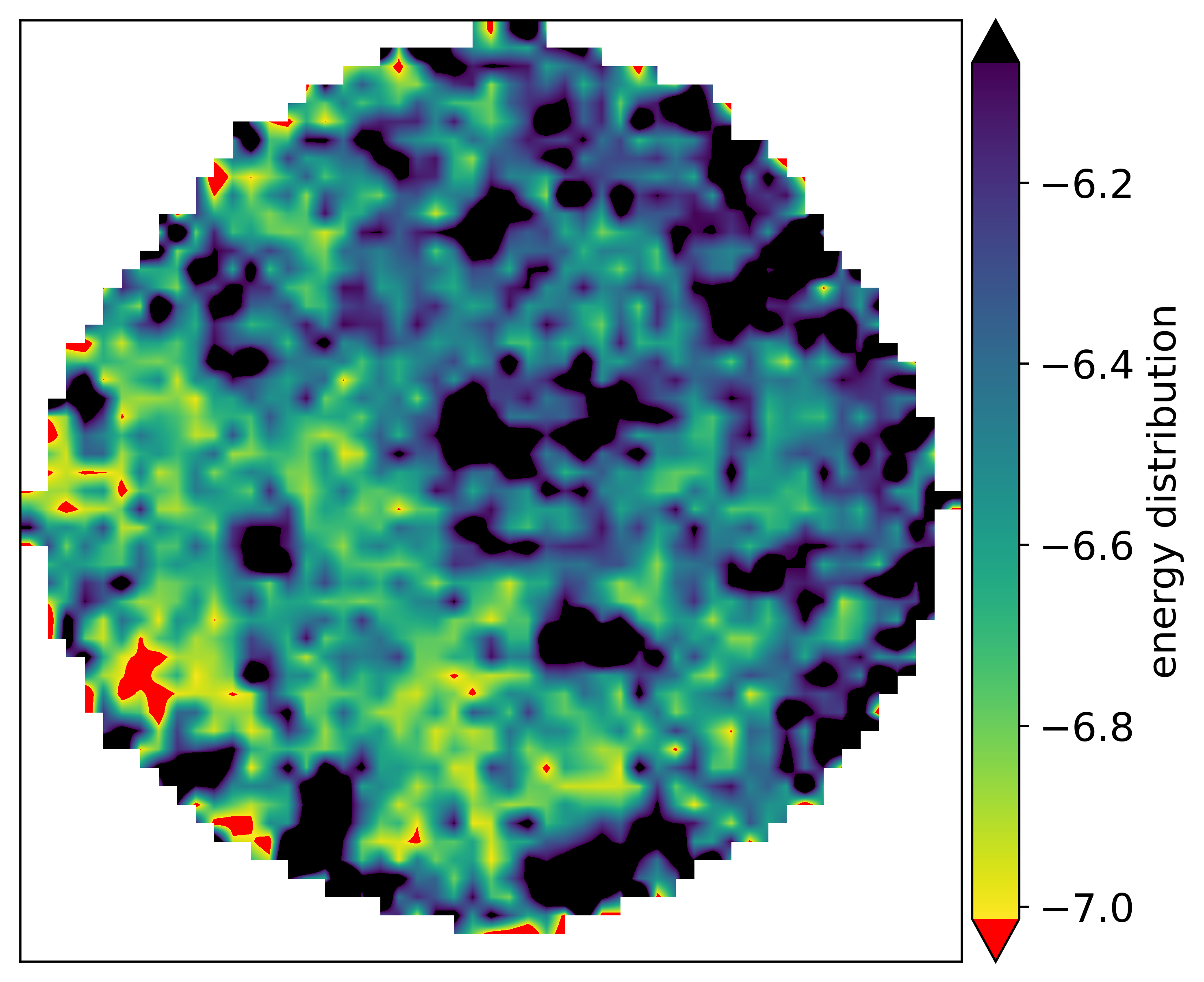}
        \caption{t-SNE$_{12}$ QC$_{LiH}$}
        \label{fig:t-SNE-12}
    \end{subfigure}

    % Thrid row
    \begin{subfigure}{0.14\textwidth}
        \centering
        \includegraphics[width=\textwidth]{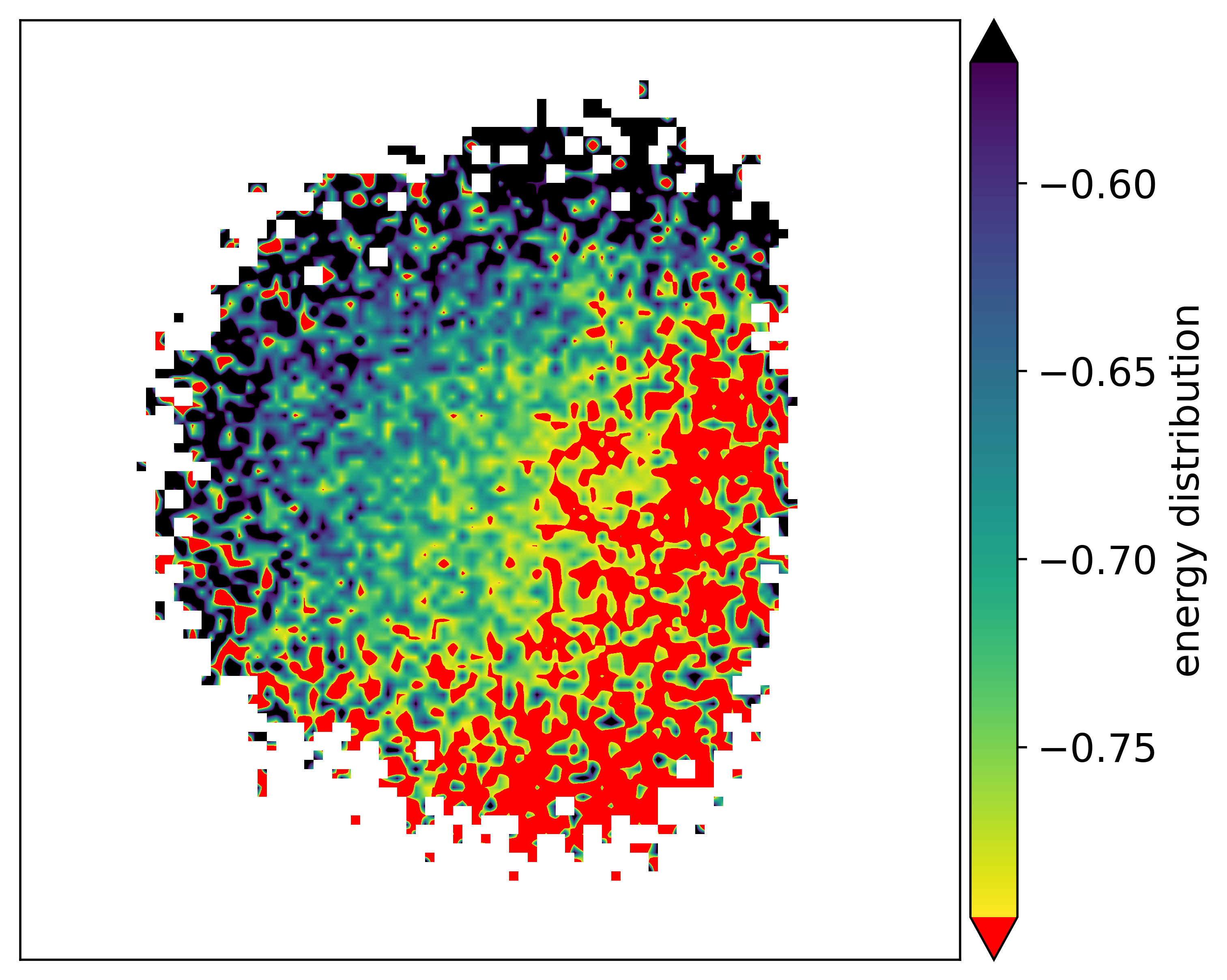}
        \caption{PCA$_4$ Q}
        \label{fig:Prev PCA-4 QC}
    \end{subfigure}
    %\hspace{0.02\textwidth}
    \begin{subfigure}{0.14\textwidth}
        \centering
        \includegraphics[width=\textwidth]{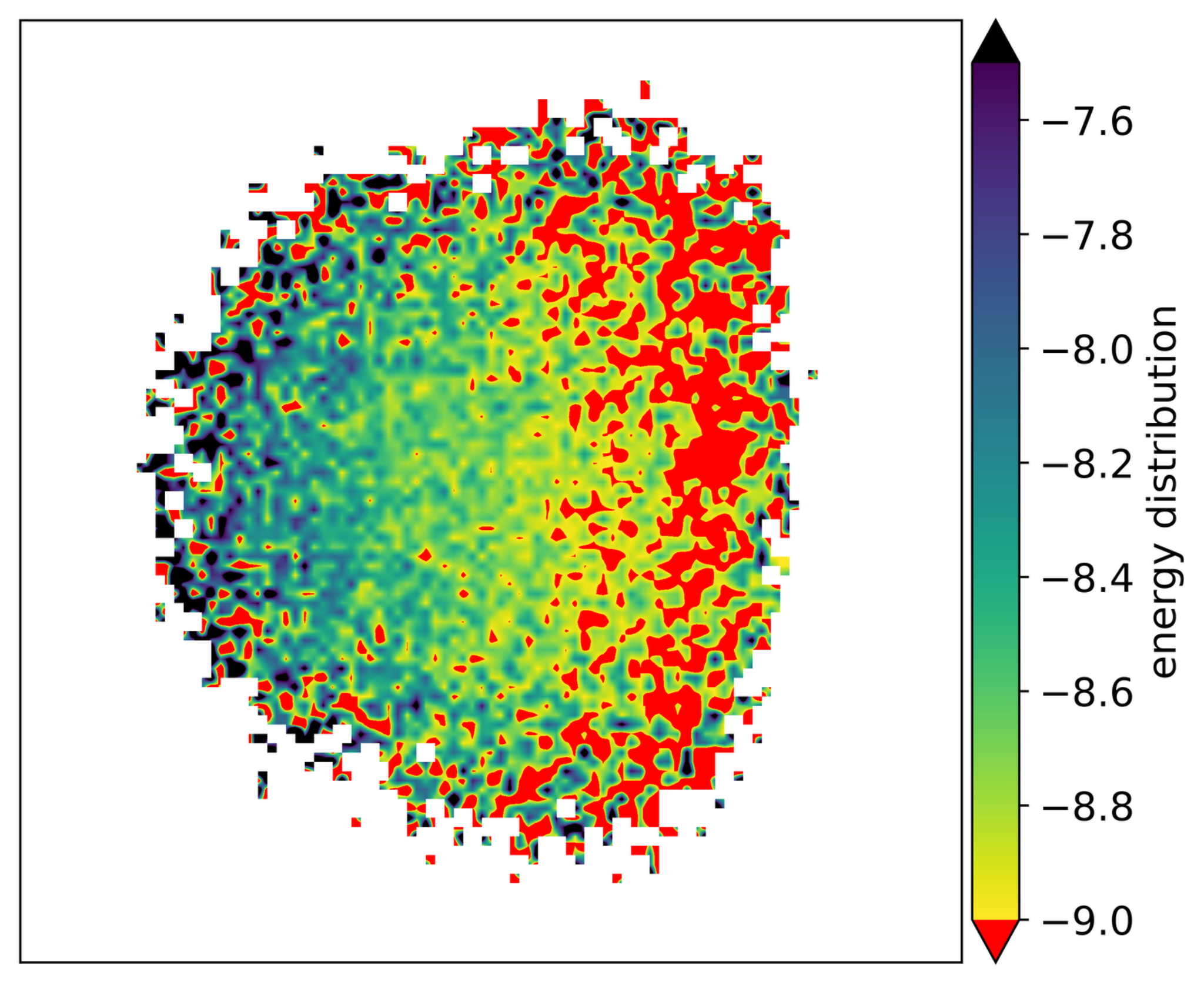}
        \caption{PCA$_4$ M}
        \label{fig:Prev PCA-4 Maxcut}
    \end{subfigure}
    \begin{subfigure}{0.14\textwidth}
        \centering
        \includegraphics[width=\textwidth]{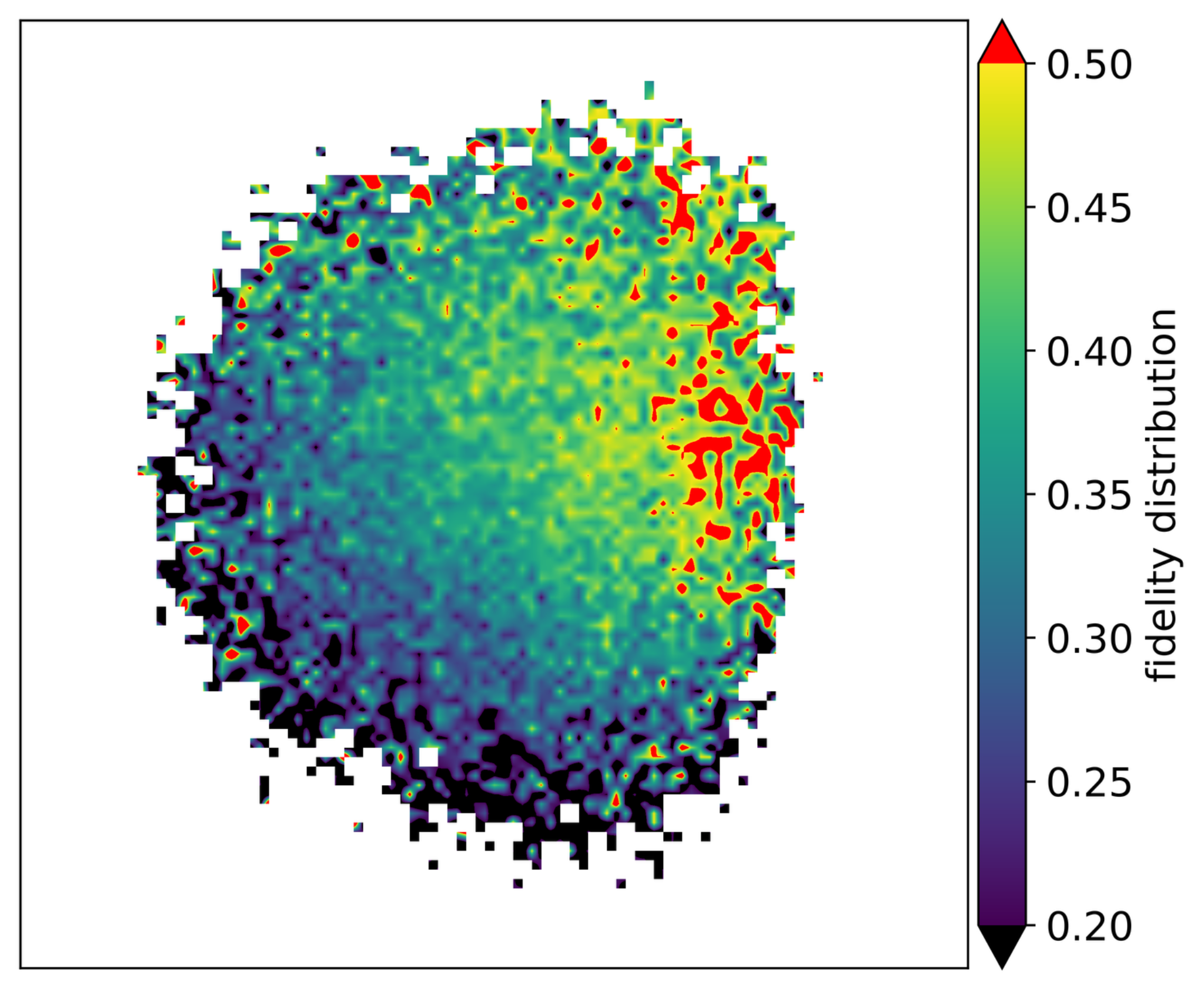}
        \caption{PCA$_4$ F}
        \label{fig:Prev PCA-4 Fidelity}
    \end{subfigure}
    \begin{subfigure}{0.14\textwidth}
        \centering
        \includegraphics[width=\textwidth]{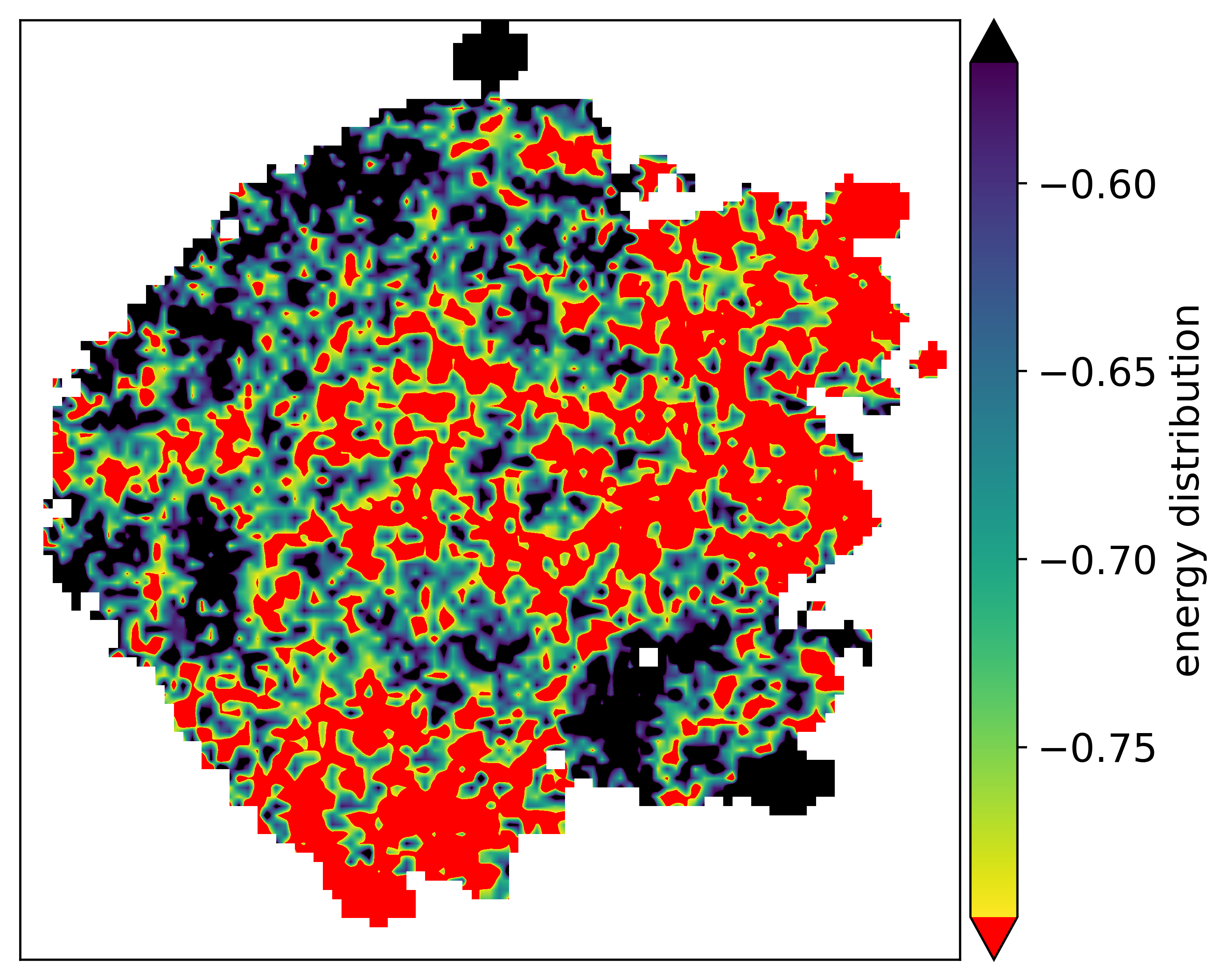}
        \caption{t-SNE$_4$ Q}
        \label{fig: Prev t-SNE-4 vqe}
    \end{subfigure}
    \begin{subfigure}{0.14\textwidth}
        \centering
        \includegraphics[width=\textwidth]{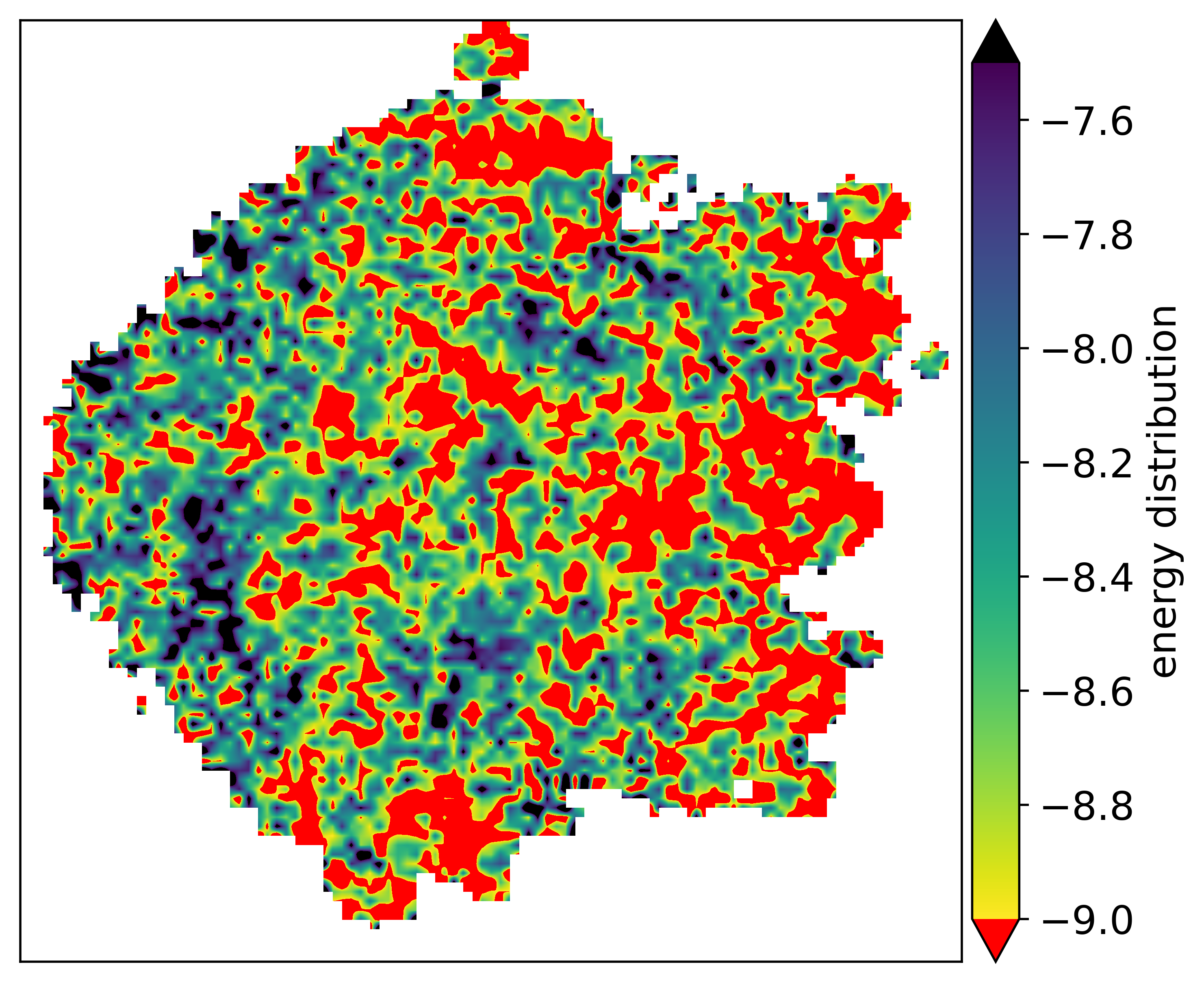}
        \caption{t-SNE$_4$ M}
        \label{fig:Prev t-SNE 4 maxcut}
    \end{subfigure}
    \begin{subfigure}{0.14\textwidth}
        \centering
        \includegraphics[width=\textwidth]{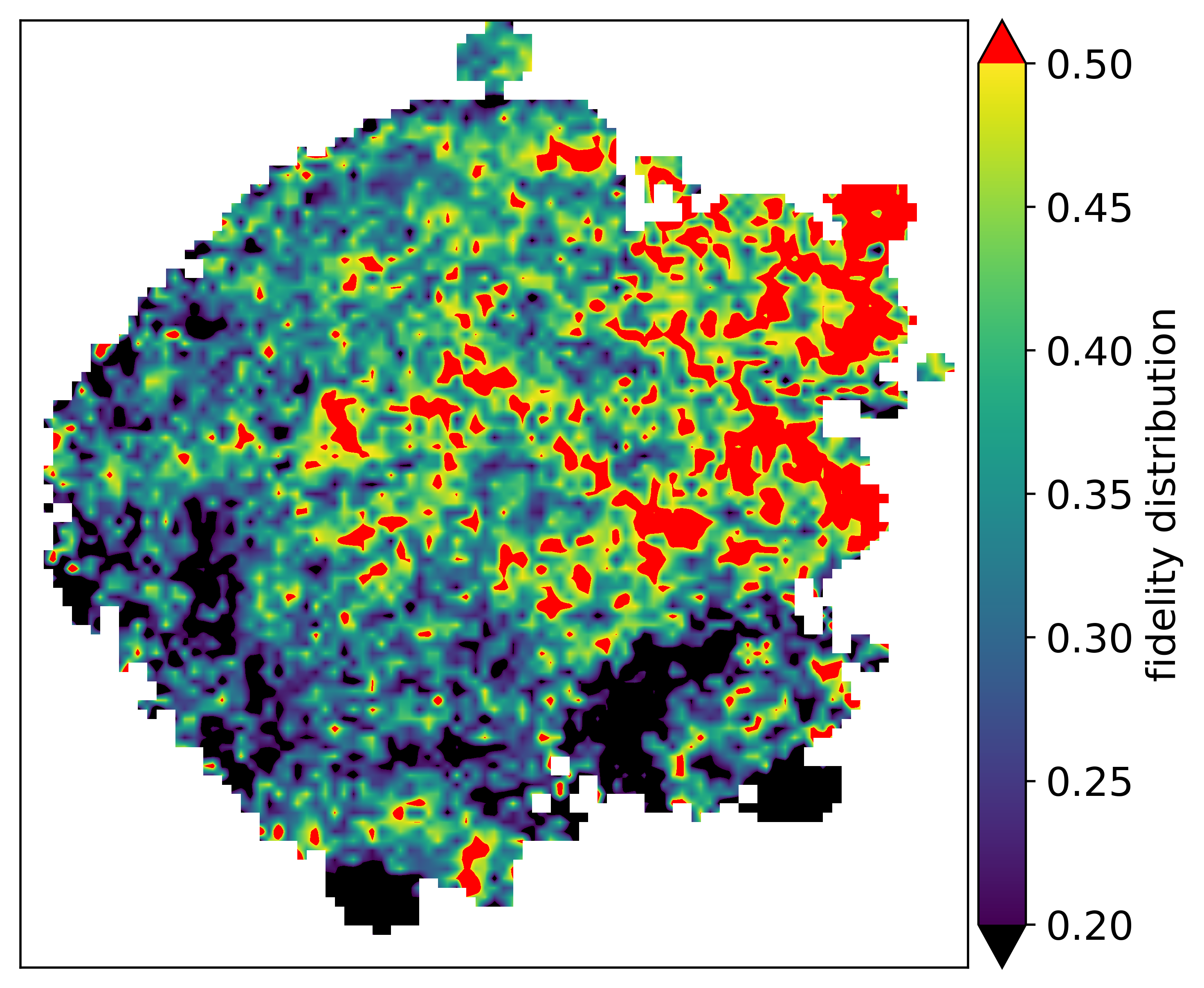}
        \caption{t-SNE$_4$ F}
        \label{fig:prev t-sne4 fidelity}
    \end{subfigure}
    
    \caption{The 2D smooth visualizations of the latent representations for the 4- and 12-qubit cases, using PCA and t-SNE. The color encoding reflects the achieved energy of 100,000 randomly generated circuits. These latent representations are introduced for three QML tasks: Quantum Chemistry, Max-cut, and fidelity. The graphs illustrate the energy or fidelity distribution of the circuits, where red denotes circuits with an energy lower than $-0.80/-0.90/-7.01 , \text{Ha}$ or a fidelity higher than 0.5. QC stands for Quantum Chemistry in figures. The subfigures in the first two rows display the results of our model with KL divergence, while the subfigures at the bottom visualize the 4-qubit latent space using the existing encoding scheme $\mathcal{E}^{GSQAS}$. \newtext{Performance values (e.g., energy or fidelity) are shown only for visualization and analysis; they are not used during the unsupervised representation learning stage and do not serve as training targets for the autoencoder.}\newtext{Best circuits correspond to those achieving the highest VQA reward or objective value during the latent-space architecture search. VQA evaluations are not used during the unsupervised representation learning stage.}}
    \label{2D-Visu}
\end{figure*}

\textbf{Observation (2):} In Figure \ref{2D-Visu}, we employ two popular techniques, PCA \citep{shlens2014tutorial} and t-SNE \citep{van2008visualizing}, to visualize the high-dimensional latent representations of 4- and 12-qubit quantum machine learning (QML) applications based on our pre-trained models. The results highlight the effectiveness of our new encoding approach for unsupervised clustering and high-dimensional data visualization. The figures show that the latent representation space of quantum circuits is smooth and compact, with architectures of similar performance clustering together when the search space is limited to 4 qubits. Notably, high-performance quantum circuit architectures are concentrated on the right side of the visualizations. In particular, PCA yields exceptionally smooth and compact representations with strong clustering effects, making it easier and more efficient to conduct QAS within such a structured latent space. This provides a robust foundation for our QAS algorithms.

For the 12-qubit latent space, high-performance circuits (shown in red) are less prominent, likely due to the fact that the 100,000 circuit structures represent only a finite subset of the possibilities for 12-qubit circuit. As a result, the number of circuits that can be learned is limited. Most high-performance circuits are distributed along the left edge of the latent space, with a color gradient transitioning from dark to light as one moves from right to left.

Compared with subfigures \ref{fig:Prev PCA-4 QC}, \ref{fig:Prev PCA-4 Maxcut}, \ref{fig:Prev PCA-4 Fidelity}, \ref{fig: Prev t-SNE-4 vqe}, \ref{fig:Prev t-SNE 4 maxcut}, and \ref{fig:prev t-sne4 fidelity}, which utilize the encoding scheme $\mathcal{E}^{GSQAS}$ and show more loosely distributed red points, our new encoding results in a more concentrated and smoother latent representation, as demonstrated in subfigures \ref{fig:PCA-4}, \ref{fig:PCA-4-Max-cut} and \ref{fig:PCA-4-Fidelity}.

\subsection{Quantum Architecture Search (QAS) Performance}
\label{qas_performance}
\begin{figure*}[ht]

\begin{center}
    \begin{subfigure}[b]{0.49\textwidth}
    \includegraphics[width=0.49\textwidth]{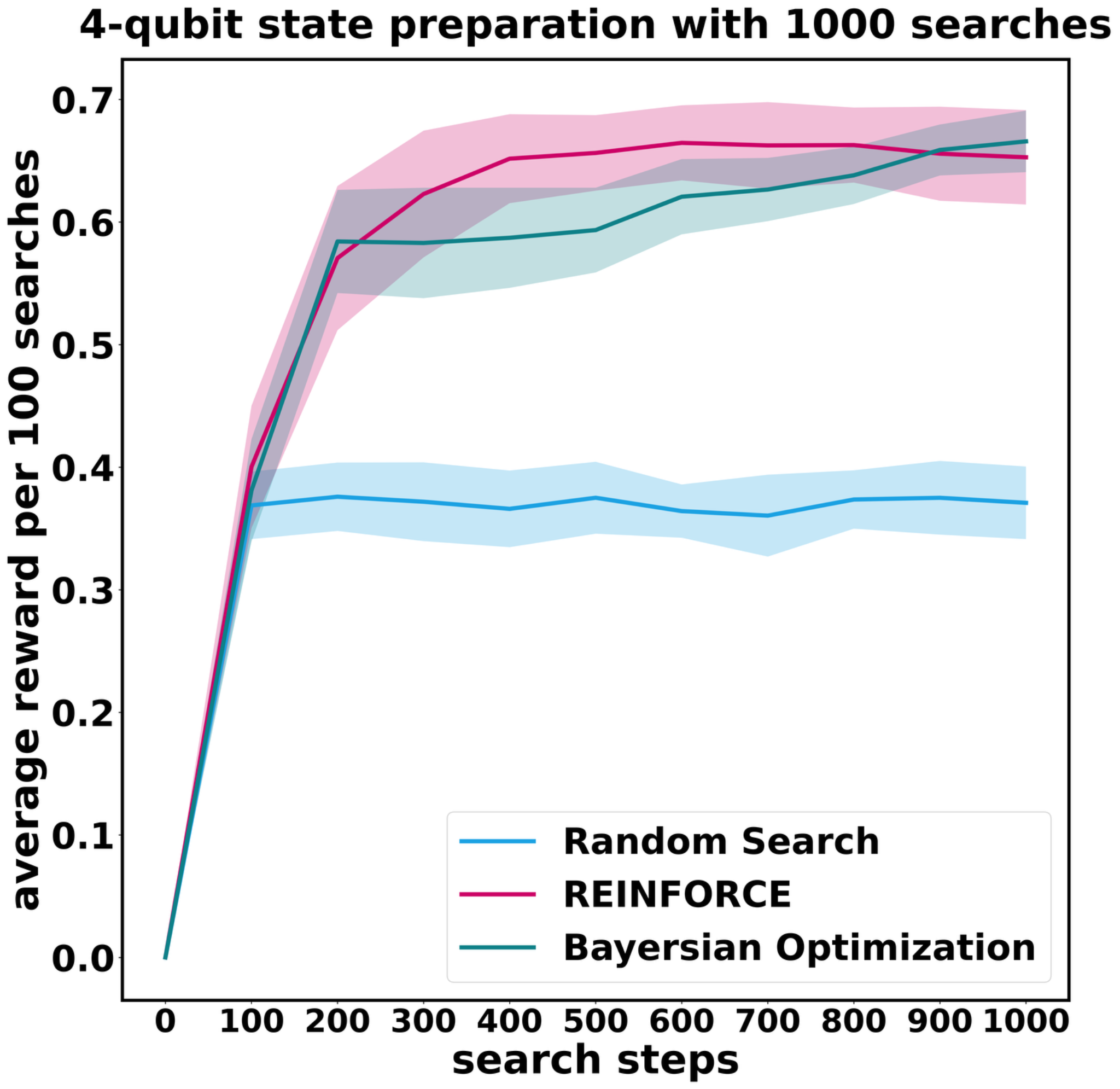} %
    \includegraphics[width=0.49\textwidth]{images/new_8-qubits-fidelity_avg_reward_per_100_with_var_filling.png} %
    \caption{State preparation}
    \label{state_preparation}
    \end{subfigure}
    \hspace{0.01cm}
    \begin{subfigure}[b]{0.49\textwidth}
    \includegraphics[width=0.48\textwidth]{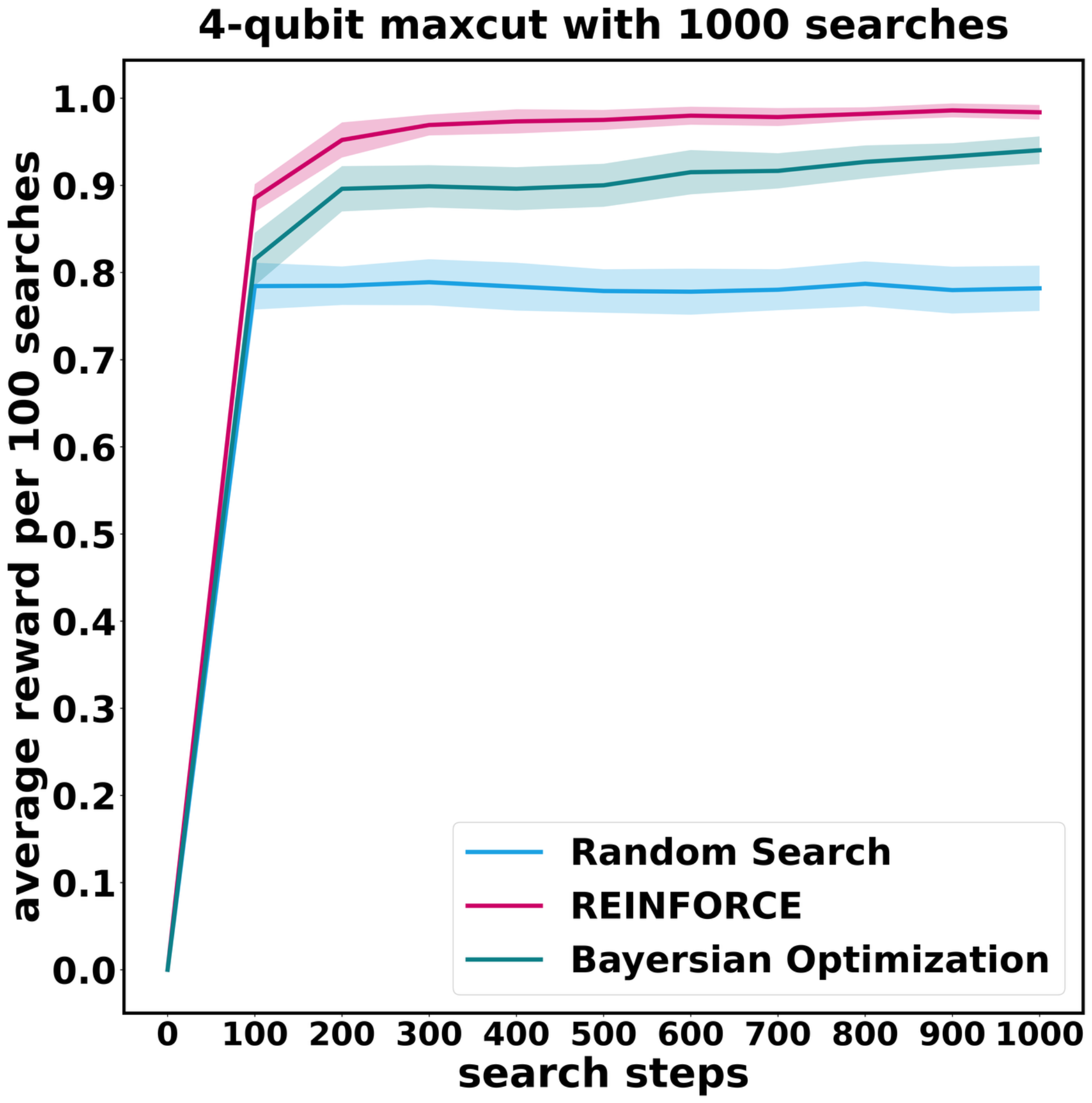}
    \includegraphics[width=0.48\textwidth]{images/new_8-qubits-maxcut_avg_reward_per_100_with_var_filling.png}
    \caption{Max-cut}
    \label{max-cut}
    \end{subfigure}
    \hspace{0.01cm}
    \begin{subfigure}[b]{0.49\textwidth}
    \includegraphics[width=0.49\textwidth]{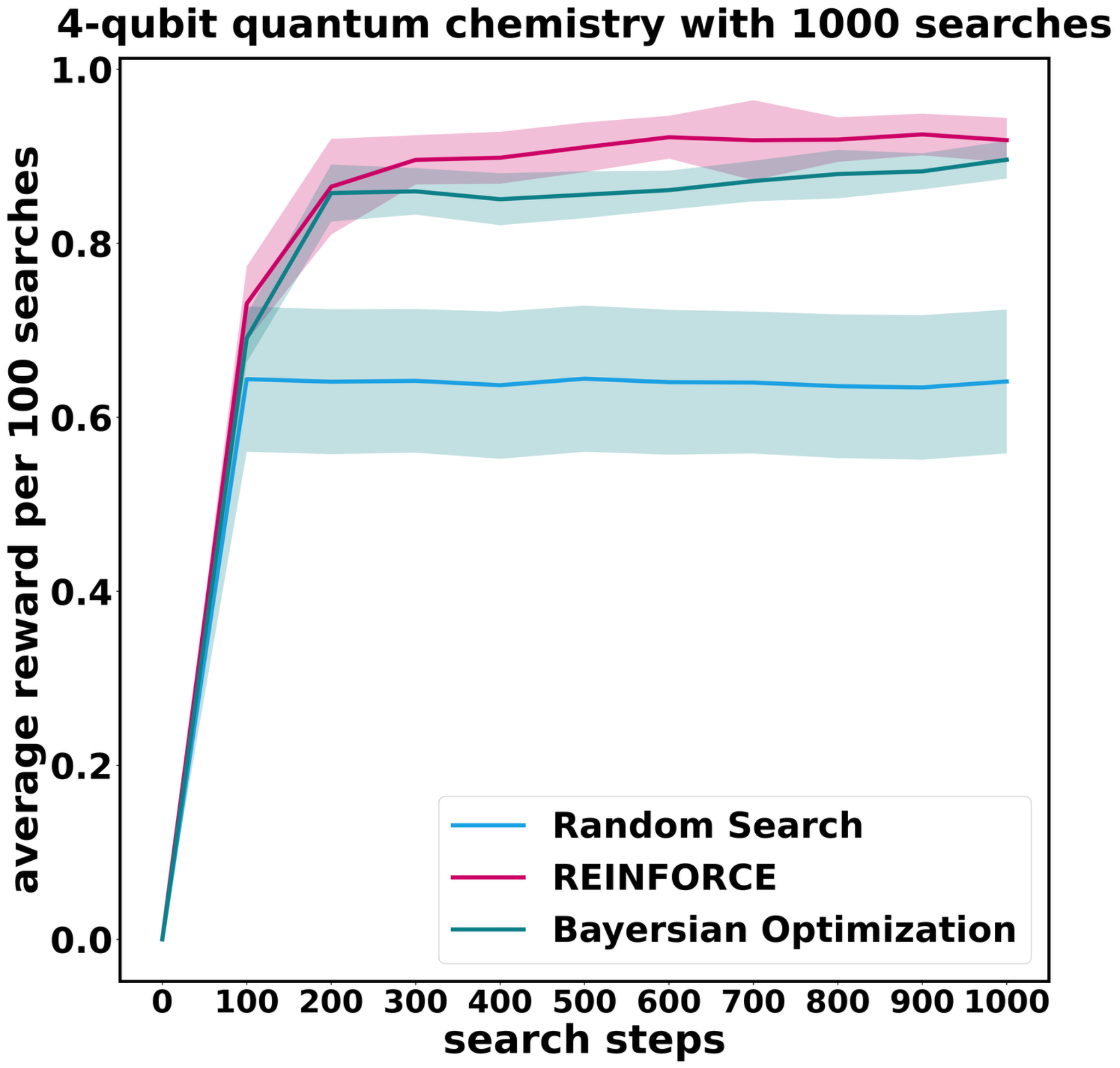}
    \includegraphics[width=0.49\textwidth]{images/new_8-qubits-vqe_avg_reward_per_100_with_var_filling.png}
    \caption{Quantum chemistry} 
    \label{quantum_chemistry}
    \end{subfigure}
    \caption{Average rewards from the six sets of experiments. In subfigures (a), (b), and (c), the left panels show results from the 4-qubit experiments, while the right panels show results from the 8-qubit experiments. Each plot presents the average reward across 50 independent runs (each with different random seeds) given 1000 search queries. The shaded areas in the plots represent the standard deviation of the average rewards.}
    \label{average_reward}
\end{center}
\end{figure*}

\textbf{Observation (1):} In Figure \ref{average_reward}, we present the average reward per 100 searches for each experiment. The results show that both the REINFORCE and BO methods effectively learn to navigate the latent representation, leading to noticeable improvements in average reward during the early stages. In contrast, Random Search fails to achieve similar improvements. Furthermore, although the plots indicate slightly higher variance in the average reward for the REINFORCE and BO methods compared to Random Search, their overall average reward is significantly higher than that of Random Search.

\begin{figure*}[htbp!]

    \centering
    % Left side (a)
    \begin{subfigure}[b]{0.49\textwidth}
        \centering
        \begin{minipage}[b]{0.6\textwidth}
            \centering
            \includegraphics[width=0.95\textwidth]{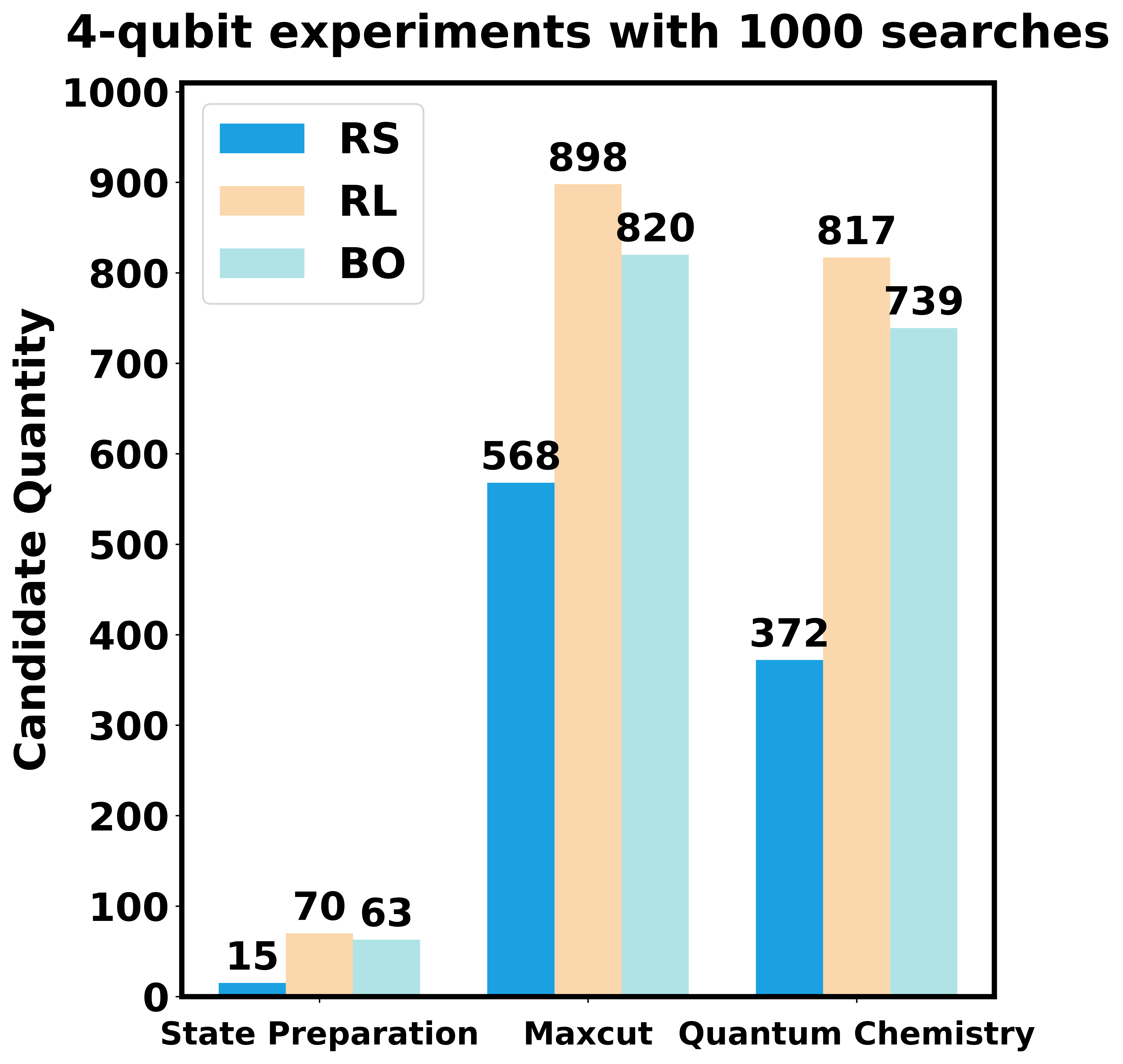}
        \end{minipage}
        \hspace{-1em}
        \begin{minipage}[b]{0.39\textwidth}
            \centering
            \includegraphics[width=\textwidth]{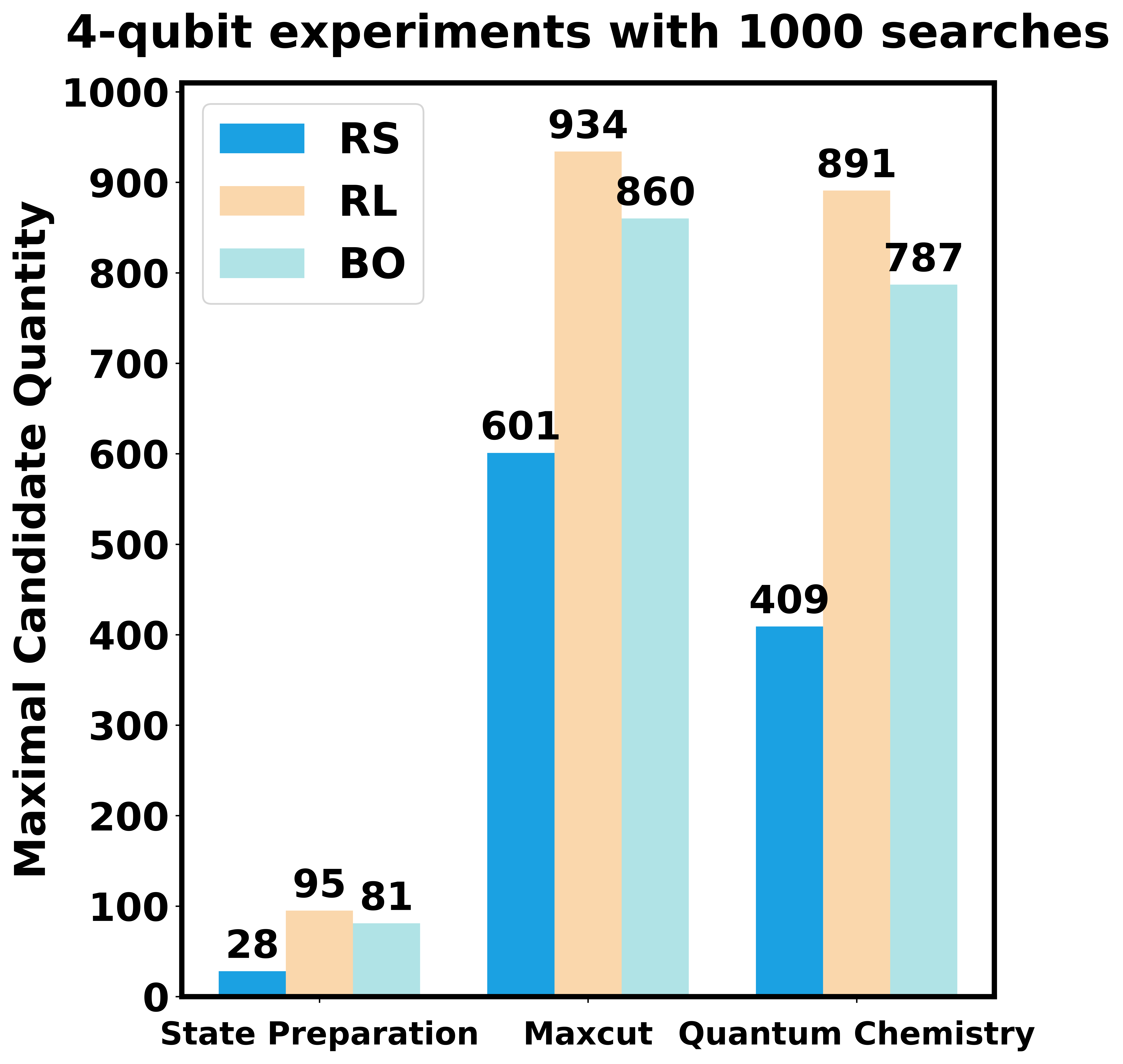}
            \includegraphics[width=\textwidth]{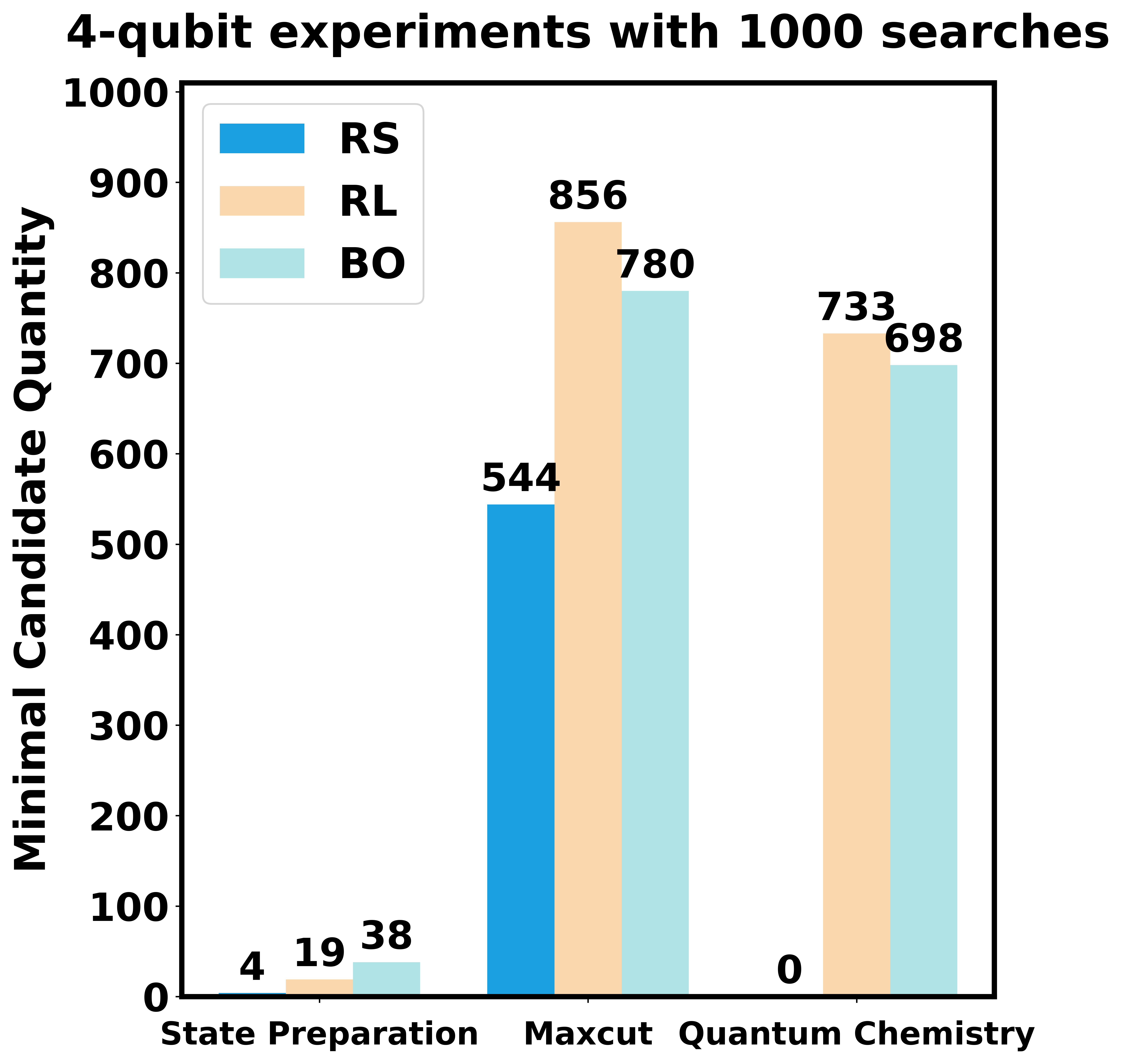}  % replace with actual inset image file path
        \end{minipage}
    \caption{4-qubit experiments}
    \label{4-qubits_candidates}
    \end{subfigure}
    \hspace{-1.5em}
    %\hfill
    \begin{subfigure}[b]{0.49\textwidth}
        \centering
        % Inset figures for (b)
        \begin{minipage}[b]{0.6\textwidth}
            \centering
            \includegraphics[width=0.95\textwidth]{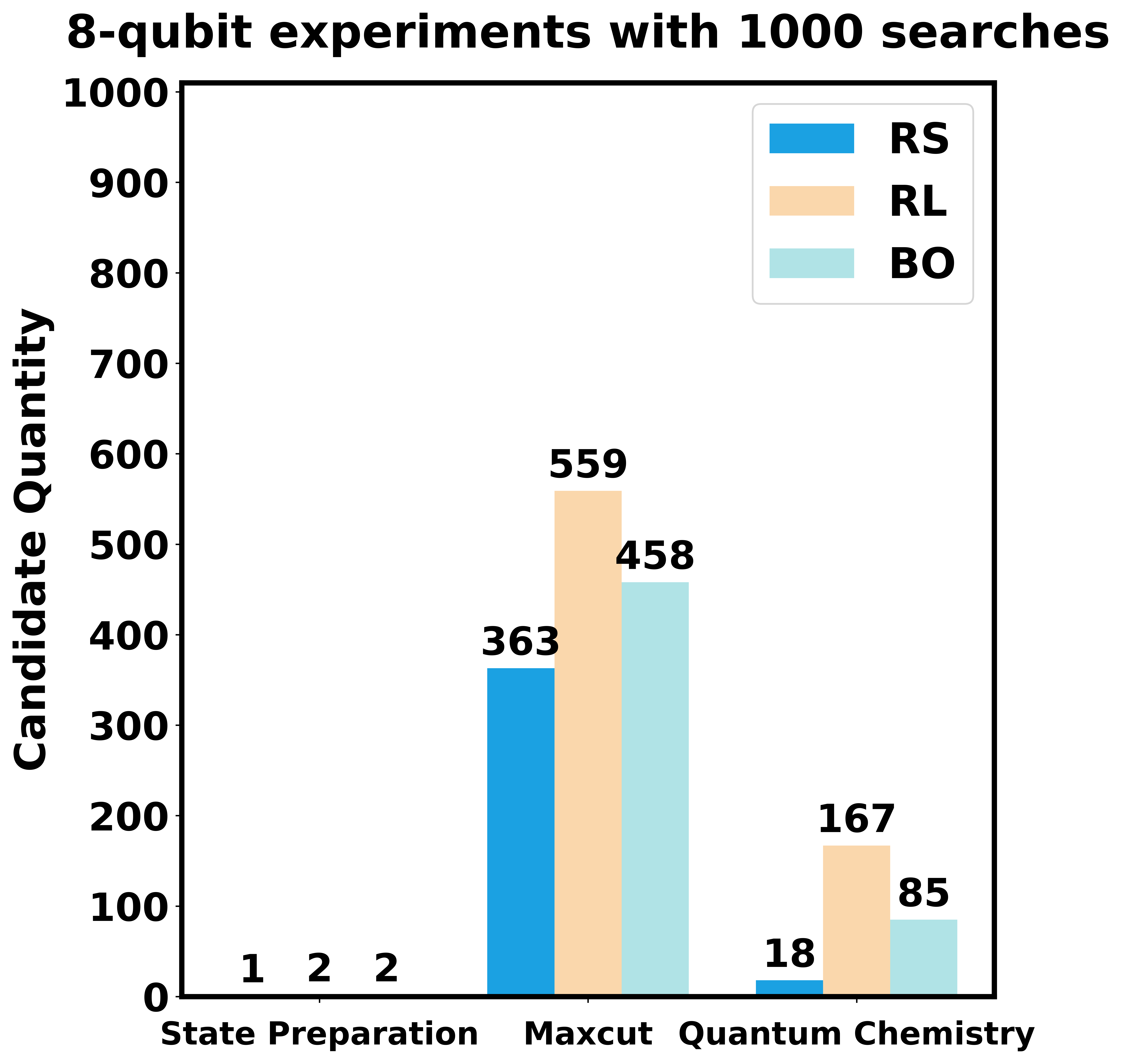}
        \end{minipage}
        \hspace{-1em}
        \begin{minipage}[b]{0.39\textwidth}
            \centering
            \includegraphics[width=\textwidth]{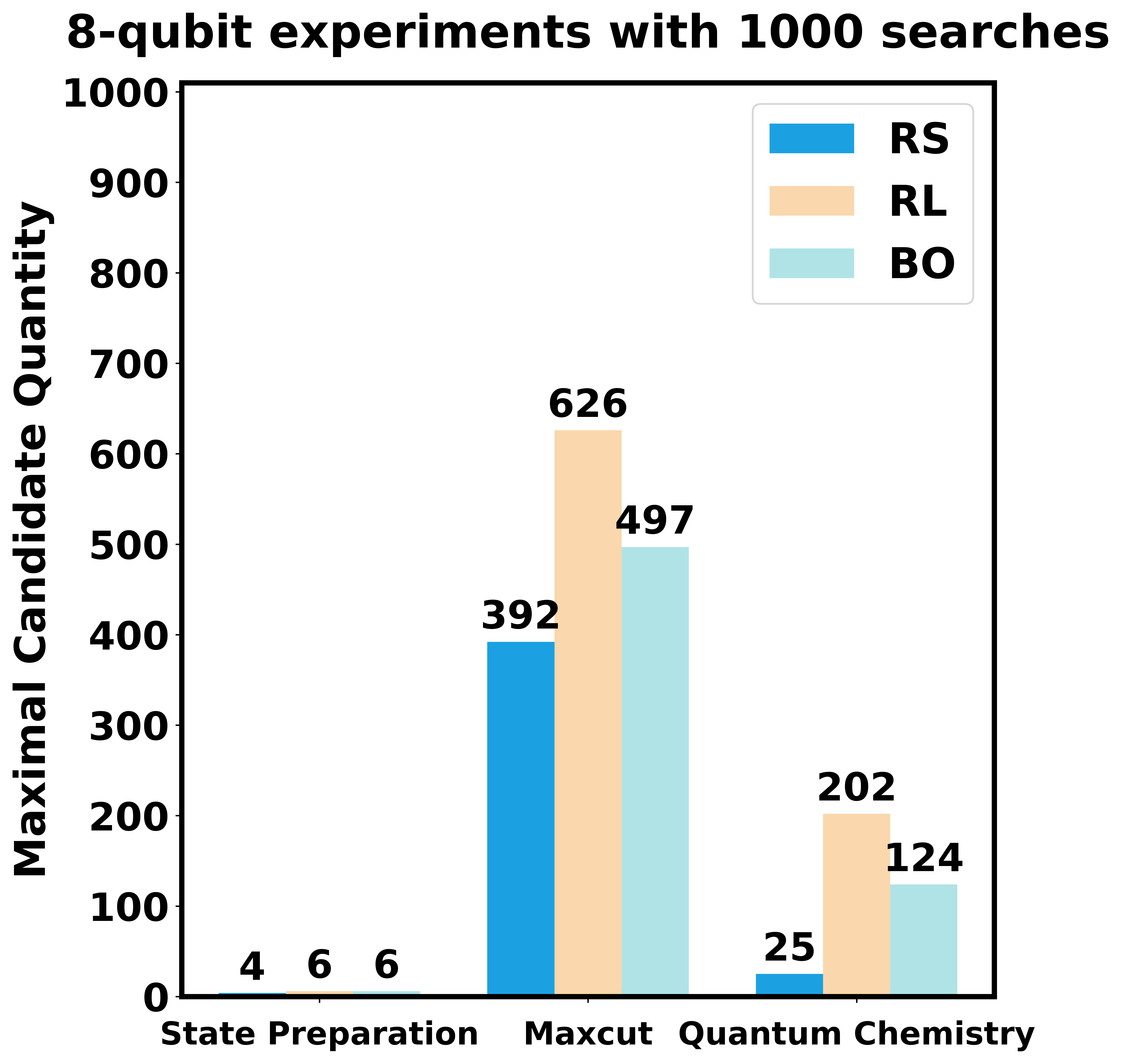}
            \includegraphics[width=\textwidth]{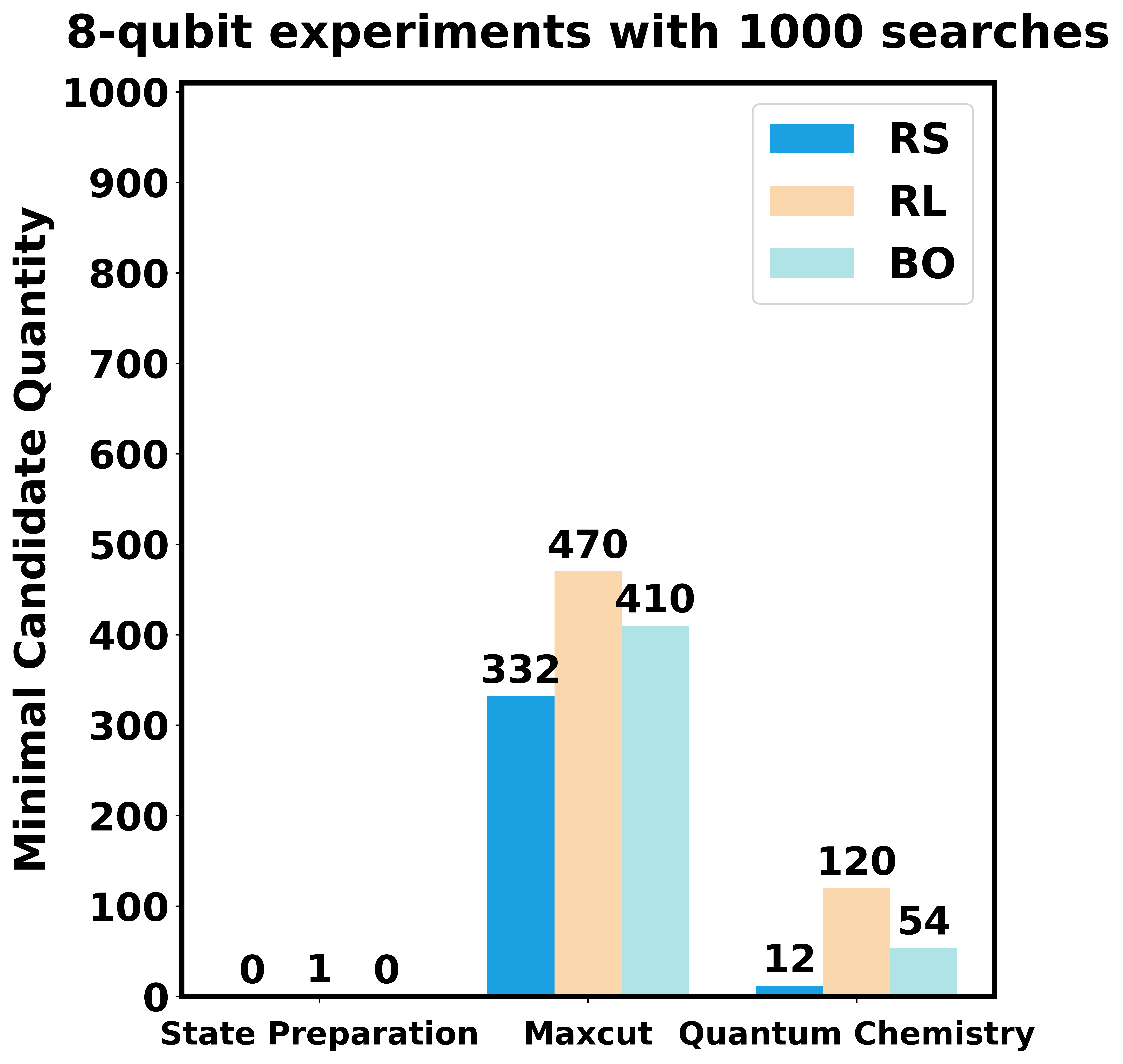}  % replace with actual inset image file path
        \end{minipage}
    \caption{8-qubit experiments}
    \label{8-qubits_candidates}
    \end{subfigure}
    \caption{The candidate quantities for the 4-qubit and 8-qubit applications. RS, RL, and BO refer to Random Search, REINFORCE, and Bayesian Optimization, respectively. The reward threshold for all 4-qubit experiments is 0.95, while for the more complex 8-qubit experiments, the thresholds are softer: 0.75 for state preparation, 0.925 for max-cut, and 0.95 for quantum chemistry. Each experiment is performed with 1000 queries, meaning only 1000 samples are drawn from a search space of 100,000 circuits. Additionally, the left-hand side of subfigures (a) and (b) shows the average results over 50 runs (with different random seeds), while the right-hand side shows the maximum and minimum candidate quantities across the 50 runs.}
    \label{candidates}
\end{figure*}

\textbf{Observation (2):} In Figure \ref{candidates}, we illustrate the number of candidate circuits found to achieve a preset threshold after performing 1000 searches using the three search methods. The results show that the 8-qubit experiments are more complex, resulting in fewer circuits meeting the requirements within the search space. Additionally, within a limited number of search iterations, both the REINFORCE and BO methods are able to discover a greater number of candidate circuits that meet the threshold, even in the worst case, i.e., when comparing the minimal number of candidates. Notably, their performance significantly surpasses that of the Random Search method, especially REINFORCE, despite the fact that the difference between the minimal and maximal number of candidates indicates that REINFORCE is more sensitive to the initial conditions compared to the other two approaches. These findings highlight the clear improvements and advantages introduced by QAS based on the latent representation, which enables the efficient discovery of numerous high-performance candidate circuits while reducing the number of searches required.

\begin{table*}[ht!]
\centering % centering table
\scriptsize
\begin{tabular}{l | l l l lllll} % creating 10 columns
\hline\hline
\addlinespace[0.5ex]
 Method & $Task$ & $F_{thr}$ & $N_{lbl}$ & $N_{rest}$ & $N_{>0.95}$ & $N_{eval}$ & $N_{QAS}$ & $N_{QAS}/N_{eval}$
\\ [0.5ex]  
\hline
\addlinespace[0.5ex]
% Entering 1st row  
\multirow{3}{*}{GNN$^{URL}$} & \text{Fidelity} & 0.5 & 1000 & 21683 & 780 & 2000 & 36 & 0.0180 \\[0.5ex] 
& \text{Max-Cut} & 0.9 & 1000 & 45960 & 35967 & 2000 & 783 & 0.3915 \\[0.5ex]
& \text{QC-4}$_{H_2}$ & 0.8 & 1000 & 65598 & 18476 & 2000 & 278 & 0.1390 \\[0.5ex]
% Entering 2nd row  
\multirow{3}{*}{GSQAS$^{URL}$} & \text{Fidelity} & 0.5 & 1000 & 21014 & 768 & 2000 & 37 & 0.0185 \\[0.5ex]
& \text{Max-Cut} & 0.9 & 1000 & 43027 & 33686 & 2000 & 785 & 0.3925 \\[0.5ex]
& \text{QC-4}$_{H_2}$ & 0.8 & 1000 & 30269 & 19889 & 2000 & 658 & 0.3290 \\[0.5ex]
\multirow{3}{*}{Random Search} 
& \text{Fidelity} & - & 0 & 100000 & 1606 & 1000 & 15 & 0.0150 \\[0.5ex]
& \text{Max-Cut} & - & 0 & 100000 & 57116 & 1000 & 568 & 0.5680 \\[0.5ex]
& \text{QC-4}$_{H_2}$ & - & 0 & 100000 & 37799 & 1000 & 371 & 0.3710 \\[0.5ex]
\hline
\addlinespace[0.5ex]
\multirow{3}{*}{QAS$^{URL}_{RL(BO)}$} & \text{Fidelity} & - & 0 & 100000 & \textbf{1606} & \textbf{1000} & \textbf{69}(63) & \textbf{0.0690}(0.0630) \\[0.5ex]
& \text{Max-Cut} & - & 0 & 100000 & \textbf{57116} & \textbf{1000} & \textbf{898}(820) & \textbf{0.8980}(0.8200) \\[0.5ex]
& \text{QC-4}$_{H_2}$ & - & 0 & 100000 & \textbf{37799} & \textbf{1000} & \textbf{817}(739) & \textbf{0.8170}(0.7390) \\[0.5ex]
\hline % inserts single-line 
\end{tabular}
\caption{Compare the QAS performance of different QAS methods for the 4-qubit tasks. URL denotes unsupervised representation learning, $F_{thr}$ is the threshold to filter poor-performance architectures, $N_{lbl}$, $N_{rest}$ and $N_{>0.95}$ refer to the number of required labeled circuits, rest circuits after filtering and the circuits that achieve the performance higher than 0.95 in the rest circuits respectively. $N_{eval}$ represents the number of evaluated circuits, i.e. the sum of the number of labeled and sampled circuits, $N_{QAS}$ is the number of searched candidates in average of 50 runs.}
\label{comparison1}
\end{table*}

\begin{table*}[ht!]
\centering % centering table
\scriptsize  % Change to \tiny, \scriptsize, \footnotesize, \small, \normalsize, \large, \Large, \LARGE, or \huge as needed
\begin{tabular}{l | l llll}
\hline\hline
\addlinespace[0.5ex]
 Method & Encoding $\mathcal{E}$ & $N_{rest}$ & $N_{eval}$ & $N_{QAS}$ & $N_{QAS}/N_{eval}$
\\ [0.5ex]  
\hline
\addlinespace[0.5ex]
% Entering 1st row  
\multirow{2}{*}{GSQAS$_{4}$}
& \text{GSQAS} & 25996 & 2000 & 625 & 0.3125 \\[0.5ex]
& \text{Ours} & 30269 & 2000 & \textbf{658} & \textbf{0.3290} \\[0.5ex]
\hline
\addlinespace[0.5ex]
\multirow{2}{*}{GSQAS$_{12}$}
& \text{GSQAS} & 60088 & 2000 & \textbf{283} & \textbf{0.1415} \\[0.5ex]
& \text{Ours} & 60565 & 2000 & 276 & 0.1380 \\[0.5ex]
\hline
\addlinespace[0.5ex]
\multirow{2}{*}{QAS$_{RL-4}$} 
& \text{GSQAS} & 100000 & 1000 & 760 & 0.7600 \\[0.5ex]
& \text{Ours} & 100000 & 1000 & \textbf{817} & \textbf{0.8170} \\[0.5ex]
\hline
\addlinespace[0.5ex]
\multirow{2}{*}{QAS$_{RL-8}$} 
& \text{GSQAS} & 100000 & 1000 & 160 & 0.1600 \\[0.5ex]
& \text{Ours} & 100000 & 1000 & \textbf{167} & \textbf{0.1670} \\[0.5ex]
\hline
\addlinespace[0.5ex]
\multirow{2}{*}{QAS$_{RL-12}$} 
& \text{GSQAS} & 100000 & 1000 & \textbf{422} & \textbf{0.4220} \\[0.5ex]
& \text{Ours} & 100000 & 1000 & 392 & 0.3920\\[0.5ex]
\hline
\addlinespace[0.5ex]

% [1ex] adds vertical space  
\hline % inserts single-line 
\end{tabular}
\caption{We compare the QAS performance of different encodings using various search methods. For the 4- and 12-qubit quantum chemistry tasks, we select $H_2$ and $LiH$, respectively, while for the 8-qubit task, we use the TFIM. The results represent the average of 50 runs.}
\label{comparison-2}
\end{table*}

\textbf{Observation (3):} In Table \ref{comparison1}, we compare various QAS methods with our approach on the 4-qubit state preparation task, using a circuit space of 100,000 circuits and limiting the search to 1000 queries. GNN$^{URL}$ and GSQAS$^{URL}$ represent predictor-based methods from \cite{he2023gnn} and \cite{he2023gsqas}, respectively, both employing our pre-trained model. QAS$^{URL}_{RL(BO)}$ denotes the QAS approach with REINFORCE (BO) used in this work. The average results over 50 runs indicate that both the predictor-based methods and our approach are capable of identifying a significant number of high-performance circuits with fewer samples.
However, predictor-based methods rely on labeled circuits to train predictors, introducing uncertainty as they may inadvertently filter out well-performing architectures along with poor ones. While a higher $F_{thr}$ value filters out more low-performance circuits, increasing the proportion of good architectures in the filtered space, it also sacrifices many well-performing circuits, which can lead to improved Random Search performance but at the cost of excluding some optimal circuits.
Despite these trade-offs, our method achieves comparable performance to predictor-based methods, demonstrating higher efficiency in terms of $N_{QAS}/N_{eval}$ while requiring fewer circuit evaluations.
In Appendix \ref{Best candidate circuits}, we present the best candidate circuits acquired by each of the three methods for every experiment.
\begin{table*}[t]
\centering
\begin{tabular}{lcccc}
\toprule
\textbf{Metric} & \multicolumn{2}{c}{\textbf{4-node MaxCut}} & \multicolumn{2}{c}{\textbf{8-node MaxCut}} \\
\cmidrule(lr){2-3} \cmidrule(lr){4-5}
& Simulator & Real Device & Simulator & Real Device \\
\midrule
Probability of Optimal Bitstring & 100.0\% & 100.0\% & 100.0\% & 100.0\% \\
Average Cut Value & 4 & 4 & 12 & 12 \\
Number of Shots & 10,000 & 10,000 & 10,000 & 10,000 \\
Noise Model & Noise-free & Physical (NISQ) & Noise-free & Physical (NISQ) \\
Device & Simulator & ibm\_sherbrooke & Simulator & ibm\_sherbrooke \\
\bottomrule
\end{tabular}
\caption{Comparison between simulator and real quantum device on 4-node and 8-node MaxCut problems using the best discovered circuits. Parameters were trained on a simulator and transferred directly to the real device. Results are averaged over 10,000 shots.}
\label{tab:sim-vs-real-multi}
\end{table*}

\textbf{\textcolor{black}{Observation (4)}:} In Table \ref{comparison-2}, we present the search performance across different frameworks and encoding methods, focusing on 4-, 8-, and 12-qubit quantum chemistry tasks for comparison. In most cases, our encoding method achieves the highest search efficiency, although the performance for the 12-qubit task is slightly lower than with another encoding method. Combined with the representation learning results in Figure \ref{2D-Visu}, we observe that the search is significantly more efficient when the learned circuit representation is smooth and concentrated. For the 12-qubit experiments, the circuits used for representation learning may be insufficient to fully capture the search space, leading to representation learning failures, as shown in Figure \ref{fig:PCA-12}, and resulting in a decline in search efficiency.

\newtext{We further study the robustness of our framework with respect to the latent dimension and the depth of the GIN encoder; ablation results for both RL- and BO-based search are reported in Appendix~\ref{Ab_depth_layers}}.

\subsection{Evaluation on Real Quantum Hardware}

To evaluate the deployability of the circuits discovered by our architecture search, we perform additional experiments using IBM's real quantum device \texttt{ibm\_sherbrooke}. For both the 4-node and 8-node MaxCut tasks, we selected the best-performing circuits and trained their parameters on a noise-free simulator. Once trained, we directly executed these circuits—without any further optimization—on the real device using the same parameters.

\paragraph{Observation.}
As shown in Table~\ref{tab:sim-vs-real-multi}, despite the presence of hardware noise and decoherence in the NISQ device, both MaxCut circuits retained their optimal output performance when transferred from simulator to real hardware. The circuits achieved a 100\% probability of measuring the optimal bitstring in 10,000 repeated shots, identical to the simulator outcome. These results validate that the discovered quantum architectures not only perform well in idealized environments but also translate reliably to real-world quantum processors without requiring parameter re-tuning.

\section{Conclusion}
In this work, we focus on exploring whether unsupervised architecture representation learning can enhance QAS. By decoupling unsupervised architecture representation learning from the QAS process, we successfully eliminate the need for a large number of labeled circuits. Additionally, our proposed quantum circuit encoding scheme addresses limitations in existing representations, improving search performance through more accurate and effective embeddings. Furthermore, our framework conducts QAS without relying on a predictor by directly applying search algorithms, such as REINFORCE and Bayesian Optimization (BO), to the latent representations. We have demonstrated the effectiveness of this approach through various experiments on simulator and real quantum hardware. In our framework, the success of QAS depends on the quality of unsupervised architecture representation learning and the selection of search algorithms. Thus, we recommend further investigation into architecture representation learning for QAS, as well as the development of more efficient search strategies within the latent representation space.

\section*{Acknowledgments}
The project of this workshop paper is supported with funds from the German Federal Ministry of Education and Research in the funding program Quantum Reinforcement Learning for industrial Applications (QLindA) - under project number 13N15644 and the Federal Ministry for Economic Affairs and Climate Action in the funding program Quantum-Classical Hybrid Optimization Algorithms for Logistics and Production Line Management (QCHALLenge) - under project number 01MQ22008B. The sole responsibility for the paper’s contents lies with the authors. Portions of this manuscript were generated with the assistance of OpenAI's ChatGPT-4 model for drafting and language refinement. The authors take full responsibility for the content.

\nocite{chatgpt2024}
\bibliography{my_bib}
\bibliographystyle{quantum}

\clearpage
\FloatBarrier
\appendix

\section{Appendix}
\subsection{Circuit Generator Settings}

The predefined operation pool which defines allowed gates in circuits is important for QAS as well, because a terrible operation pool such as one with no rotation gates or no control gates cannot generate numerous quantum circuits with excellent expressibility and entanglement capability. This makes the initial quantum search space poor, so it will influence our further pre-training and QAS process. Therefore, we choose some generally used quantum gates in PQCs as our operation pool $\{\texttt{X, Y, Z, H, Rx, Ry, Rz, U3, CNOT, CZ, CY}\}$ for the circuit generator to guarantee the generality of our quantum circuit space. Other settings of the circuit generator are summarized below:
\begin{table*}[ht!]
\centering
    \begin{tabular}{p{0.2\textwidth}p{0.4\textwidth}p{0.2\textwidth}}
    \hline
     Hyperparameter & Hyperparameter explanation & Value for 4/8/12-qubit experiments\\
    \hline
    num-qubits & the number of qubits & 4/8/12\\
    num-gates & the number of gates in a circuit & 10/20/30\\
    max-depth & the maximal depth in a circuit & 5\\
    num-circuits & required the number of circuits & $10^5$\\
    \bottomrule
  \end{tabular}
\caption{Description of settings predefined for the circuit generator.}
\label{generator_setting}
\end{table*}

\subsection{Application Settings}
\label{application_setting}
\begin{figure*}[ht]
\begin{center}
    \begin{subfigure}[b]{0.475\textwidth}
    \centering
    \includegraphics[width=\textwidth]{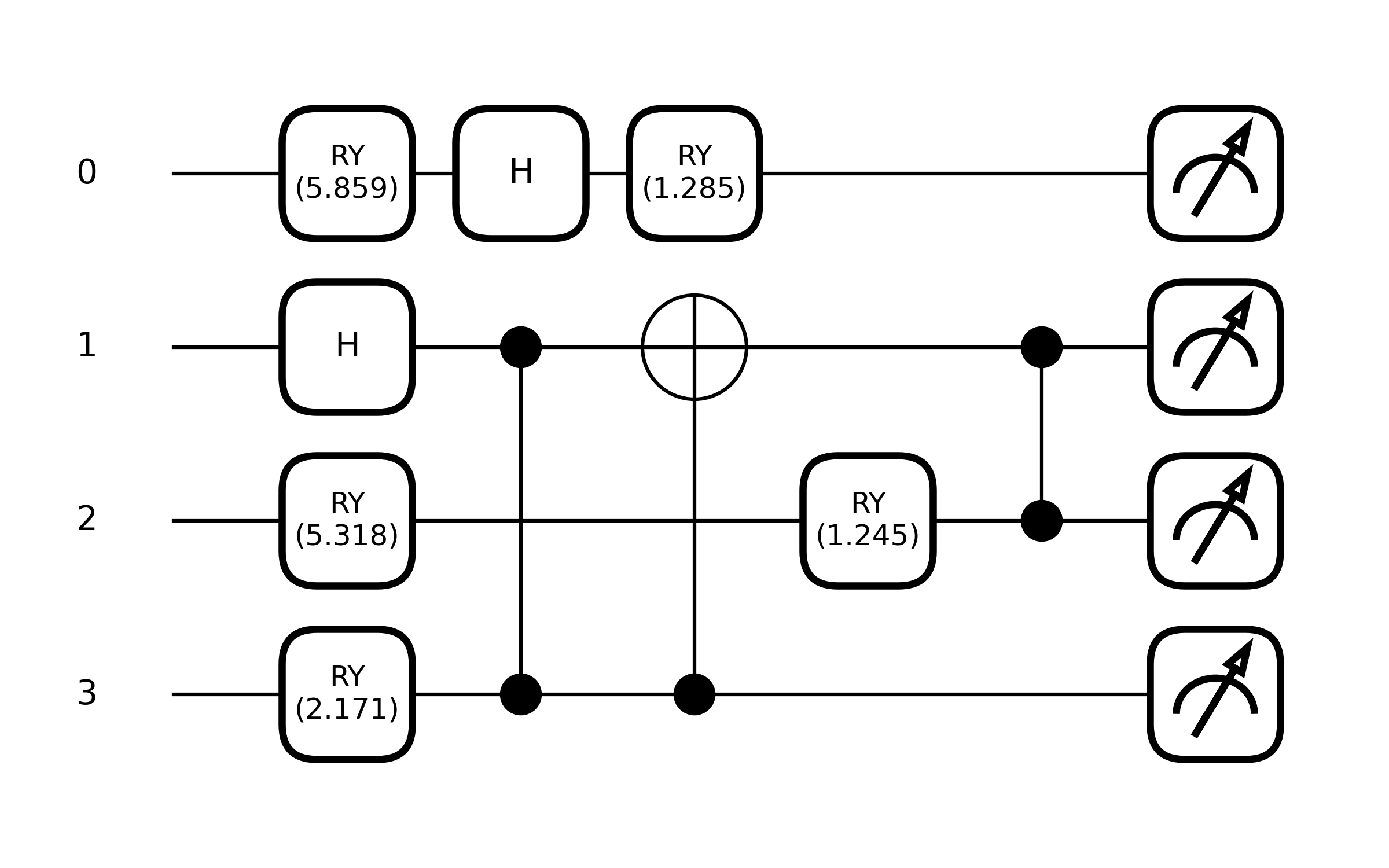} %
    \caption{The target circuit of the 4-qubit state preparation} 
    \end{subfigure}
    \hspace{0.5cm}
    \begin{subfigure}[b]{0.475\textwidth}
    \centering
    \includegraphics[width=0.75\textwidth, height=0.6\textwidth]{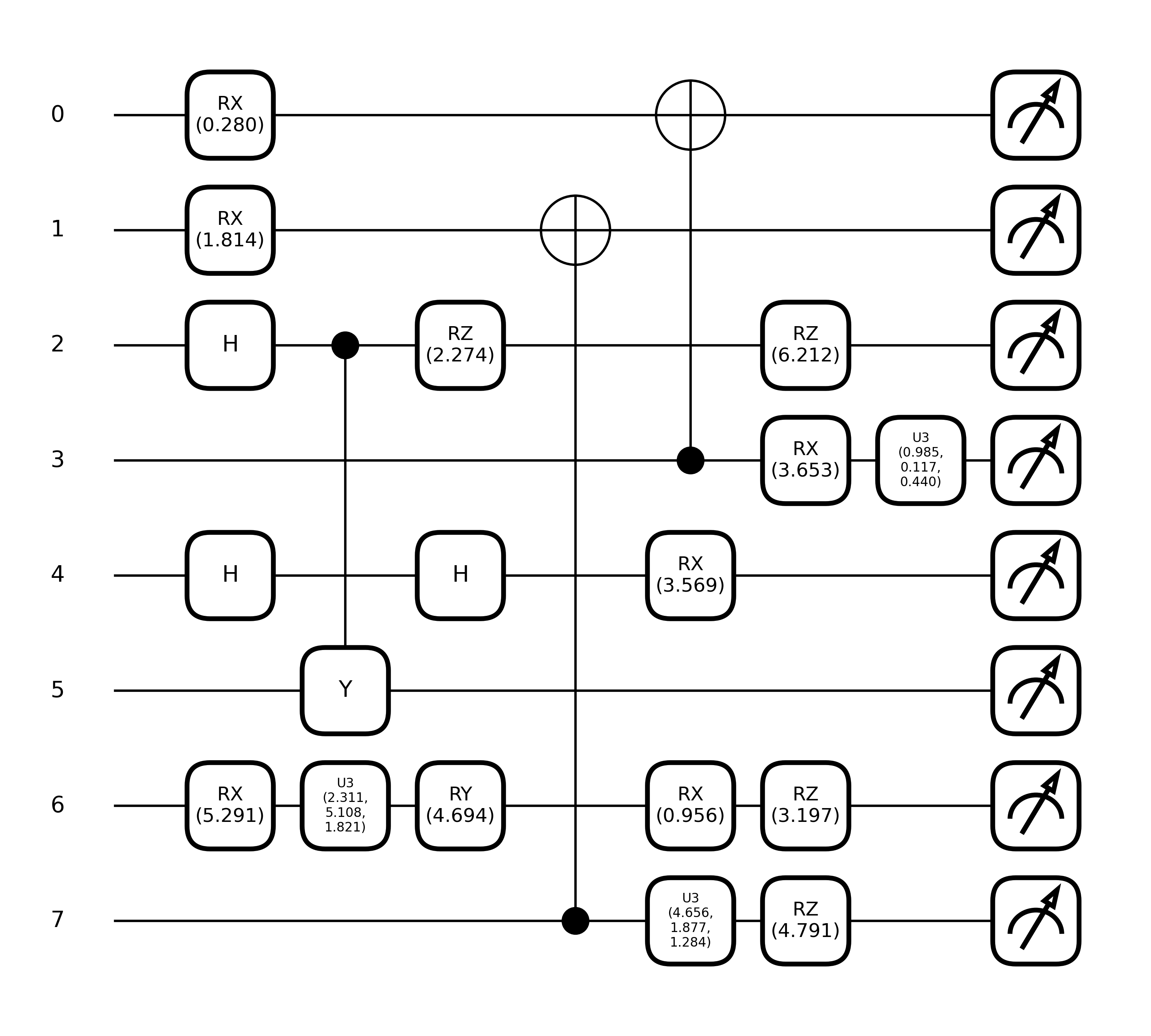}
    \caption{The target circuit of the 8-qubit state preparation} 
    \end{subfigure}
    \caption{The circuits used to generate the target states.}
    \label{fidelity_target}  
\end{center}
\end{figure*}
\paragraph{Quantum State Preparation.} In quantum information theory, fidelity \citep{liang2019quantum} is an important metric to measure the similarity of two quantum states. By introducing fidelity as the performance index, we aim to maximize the similarity of the final state density operator with a certain desired target state. We first obtain the target state by randomly generating a corresponding circuit, and then with a limited number of sample circuits, we use the search methods to search candidate circuits that can achieve a fidelity higher than a certain threshold. During the search process, the fidelity can be directly used as a normalized reward function since its range is [0, 1]. Figure~\ref{fidelity_target} shows the circuits used to generate the corresponding target states.

\paragraph{Max-cut Problems.} The max-cut problem \citep{poljak1995solving} consists of finding a decomposition of a weighted undirected graph into two parts (not necessarily equal size) such that the sum of the weights on the edges between the parts is maximum. Over these years, the max-cut problem can be efficiently solved with quantum algorithms such as QAOA \citep{villalba2021improvement} and VQE (using eigenvalues). In our work, we address the problem by deriving the Hamiltonian of the graph and using VQE to solve it. We use a simple graph with the ground state energy $-10 \ Ha$ for the 4-qubit experiment and a relatively complex graph with the ground state energy $-52 \ Ha$ in the case of the 8-qubit experiment. Furthermore, we convert the energy into a normalized reward function integral to the search process. The visual representations of these graphs are presented below:
\begin{figure*}[ht]
\begin{center}
    \begin{subfigure}[b]{0.4\textwidth}
    \centering
    \includegraphics[width=\textwidth]{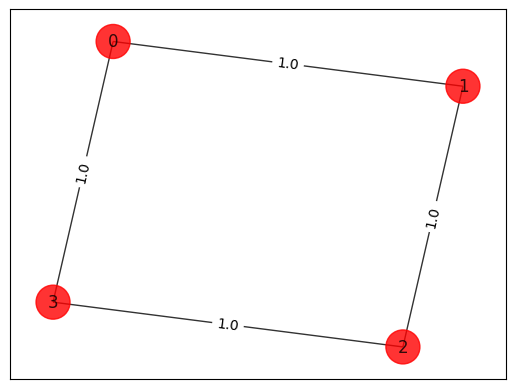} %
    \caption{The 4-qubit max-cut graph} 
    \end{subfigure}
    \hspace{1cm}
    \begin{subfigure}[b]{0.4\textwidth}
    \centering
    \includegraphics[width=\textwidth]{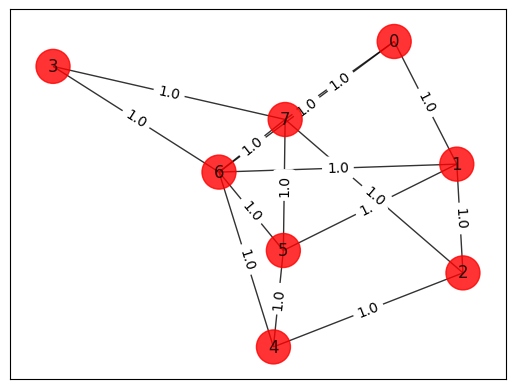}
    \caption{The 8-qubit max-cut graph} 
    \end{subfigure}
    \caption{The graphs of the experiments on max-cut problems.}
    \label{maxcut_graph}  
\end{center}
\end{figure*}

\paragraph{Quantum Chemistry.} In the field of quantum chemistry, VQE \citep{peruzzo2014variational, tilly2022variational} is a hybrid quantum algorithm for quantum chemistry, quantum simulations, and optimization problems. It is used to compute the ground state energy of a Hamiltonian based on the variational principle. For the 4- and 12-qubit quantum chemistry experiment, we use the Hamiltonian of the molecule $H_2$ and $LiH$ and its common approximate ground state energy $-1.136 \ Ha$ and $-7.88 \ Ha$ as the optimal energy. As for the 8-qubit experiment, we consider $n=8$ transverse field Ising model (TFIM) with the Hamiltonian as follows:
\begin{align}
    \mH = 
    \sum_{i=0}^7\sigma_z^i\sigma_z^{(i+1) \ mod \ 6} + \sigma_x^i\text{.}
\end{align}
We design some circuits to evaluate the ground state energy of the above Hamiltonian and get an approximate value $-10 \ Ha$ as the optimal energy. According to the approximate ground state energy, we can use our methods to search candidate circuits that can achieve the energy reaching a specific threshold. In the process of searching for candidates, the energy is normalized as a reward function with the range [0, 1] to guarantee search stability.

\subsection{\textcolor{black}{Ablation Studies}}
\label{Ab_depth_layers}

\newtext{In this subsection, we analyze the sensitivity of our method to key
hyperparameters of the unsupervised representation learning stage, namely the
latent dimension $d_z$ and the depth $L$ of the GIN encoder. We evaluate both
representation quality and downstream QAS performance using REINFORCE-based
and Bayesian Optimization–based search on the 4-qubit MaxCut task. In Table~\ref{tab:ablation_pretrain}, we observe that the reconstruction objective (VAE loss) varies with the latent dimension and GIN depth, whereas the downstream QAS performance is much less sensitive: both REINFORCE and Bayesian Optimization achieve comparable rewards across these configurations.}

\begin{table*}[t]
%\captionsetup{labelformat=blue} % <-- LOCAL, only this figure
\centering
\begin{tabular}{ccccc}
\hline
$d_z$ & $L$ & VAE loss & avg.\ reward (RL) & avg. reward (BO)\\
\hline
8  & 5 & 0.11 & $0.95 \pm 0.001$ & $0.89 \pm 0.030$\\
16 & 5 & 0.09 & $0.98 \pm 0.006$ & $0.93\pm 0.013$\\
32 & 5 & 0.14 & $0.975 \pm 0.009$ & $0.92\pm 0.017$\\
16 & 3 & 0.10 & $0.95 \pm 0.013$ & $0.93\pm 0.021$\\
16 & 7 & 0.09 & $0.97 \pm 0.010$ & $0.92 \pm 0.022$\\
\hline
\end{tabular}
\caption{Ablation on the latent dimension $d_z$ and GIN depth $L$ in the unsupervised pretraining stage on 4-qubit circuits. The VAE loss is the final-epoch training objective (reconstruction + KL), where lower is better. The avg.\ reward is obtained from the RL- / BO-based search strategy on the 4-qubit MaxCut problem and is reported as the mean $\pm$ standard deviation over 10 random seeds, where higher is better.}
\label{tab:ablation_pretrain}
\end{table*}

\subsection{Hyperparameters of Pre-training}
\label{pretraining_parameters}
Table \ref{table:appendix Description of hp} shows the hyperparameter settings of the pre-training model for 4-qubit and 8-qubit experiments.
\begin{table*}[t]
\centering
    \begin{tabular}{p{0.2\textwidth}p{0.4\textwidth}p{0.2\textwidth}}
    \hline
     Hyperparameter & Hyperparameter explanation & Value for 4/8/12-qubit experiments\\
    \hline
    bs & batch size & 32\\
    epochs & training epochs & 16\\
    dropout & decoder implicit regularization & 0.1\\
    normalize & input normalization & True\\
    input-dim & input dimension & 2+$\#$gates+$\#$qubits\\
    hidden-dim & dimension of hidden layer & 128\\
    dim & dimension of latent space & 16\\
    hops & the number of GIN layers ($L$ in eq.~\ref{eq:gin_update}) & 5 \\
    mlps & the number of MLP layers & 2 \\
    \bottomrule
  \end{tabular}
\caption{Description of hyperparameters adopted for pre-training.}
\label{table:appendix Description of hp}
\end{table*}

\subsection{Best Candidate circuits}
\label{Best candidate circuits}

\textbf{Observation (5):} In Appendix \ref{Best candidate circuits}, we present the best candidate circuits acquired by each of the three methods for every experiment. These circuits exhibit a higher likelihood of being discovered by REINFORCE and BO in contrast to Random Search. This observation underscores the superior search capabilities of REINFORCE and BO in navigating the large and diverse search space generated by our approach, which is based on a random generator derived from a fixed operation pool. Unlike conventional approaches that adhere to layer-wise circuit design baselines, our method excels in discovering circuits with fewer trainable parameters. This characteristic is of paramount importance when addressing real-world optimization challenges in QAS. In conclusion, our approach not only enhances the efficiency of candidate circuit discovery but also accommodates the distinct characteristics of various problem domains through a large and diverse search space.
\begin{figure*}[t]
\begin{center}
    \begin{subfigure}[b]{0.275\textwidth}
    \centering
    \includegraphics[width=\textwidth]{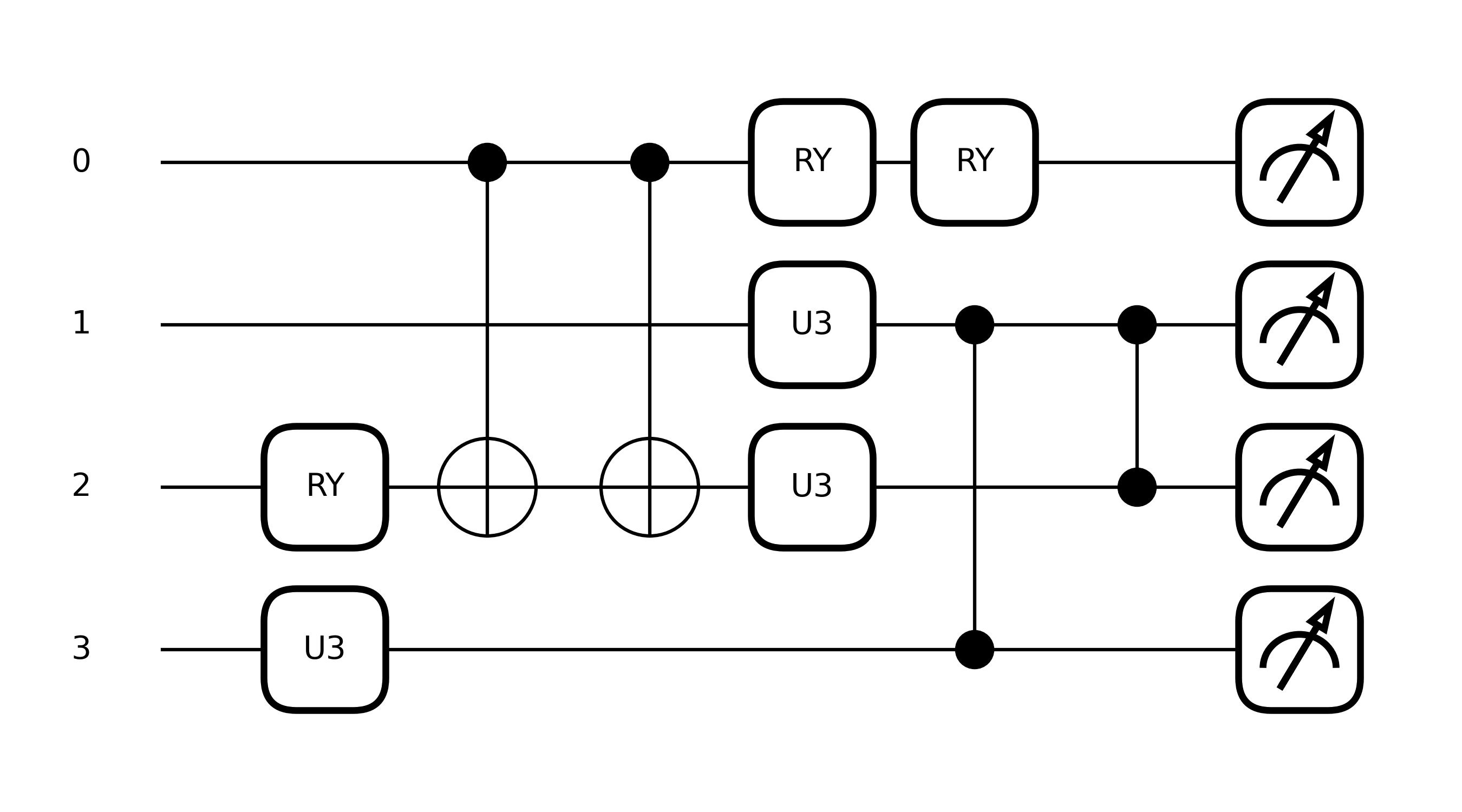}

    \caption{4-qubit state preparation}
    \end{subfigure}
    \hfill
    \begin{subfigure}[b]{0.275\textwidth}
    \centering
    \includegraphics[width=\textwidth]{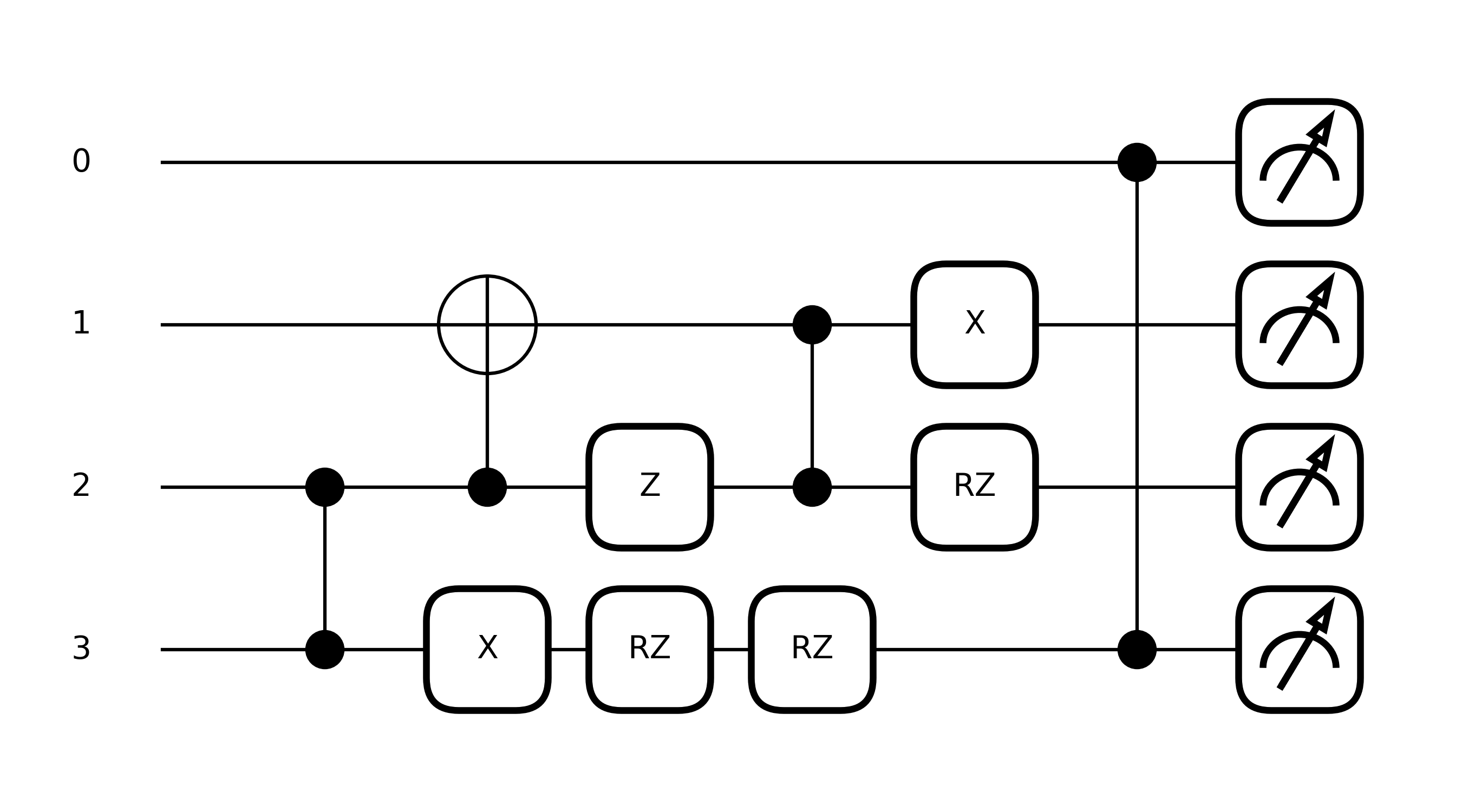}
    \caption{4-qubit max-cut}
    \end{subfigure}
    \hfill
    \begin{subfigure}[b]{0.275\textwidth}
    \centering
    \includegraphics[width=\textwidth]{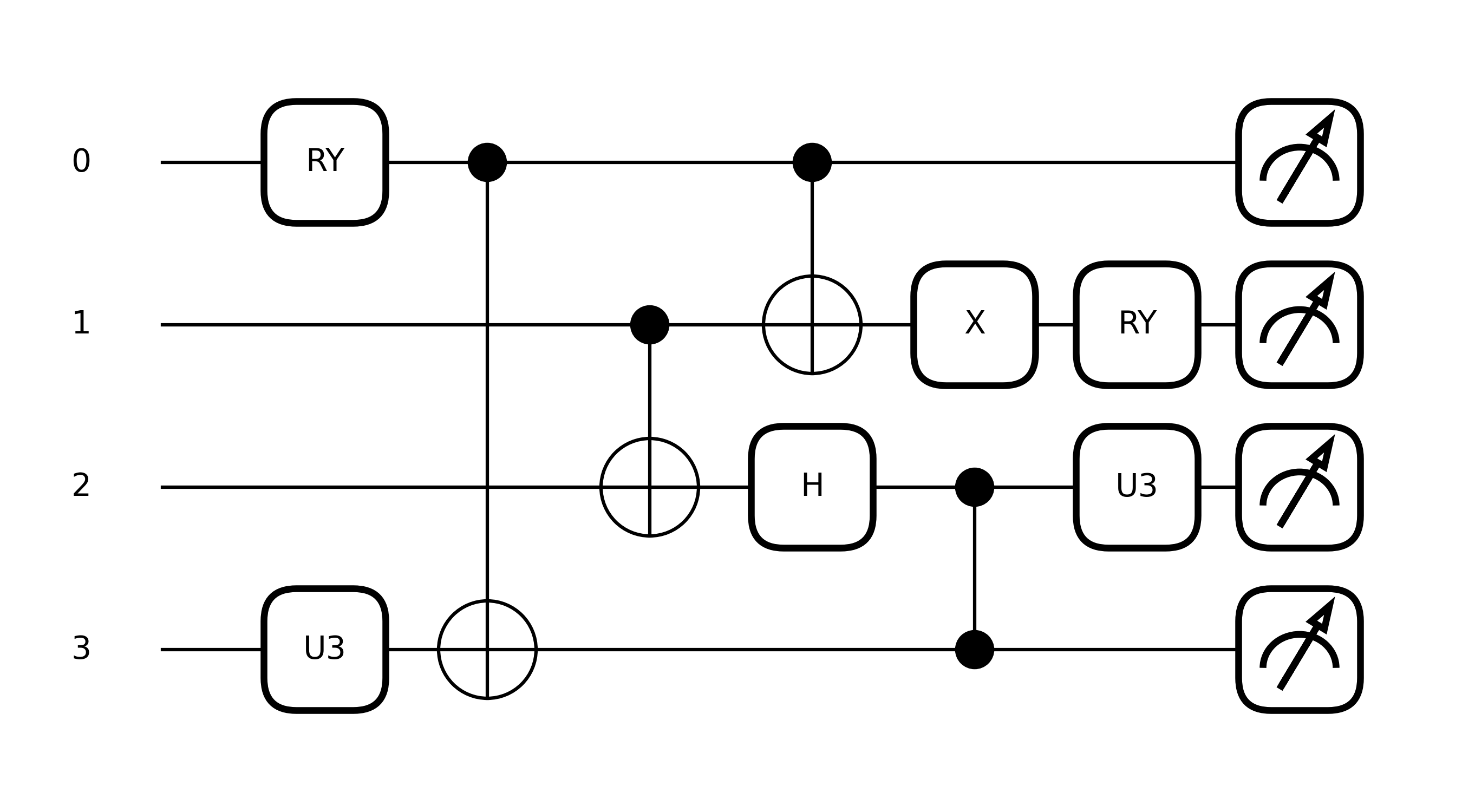}
    \caption{4-qubit quantum chemistry}
    \end{subfigure}

    \begin{subfigure}[b]{0.275\textwidth}
    \centering
    \includegraphics[width=\textwidth, height=0.75\textwidth]{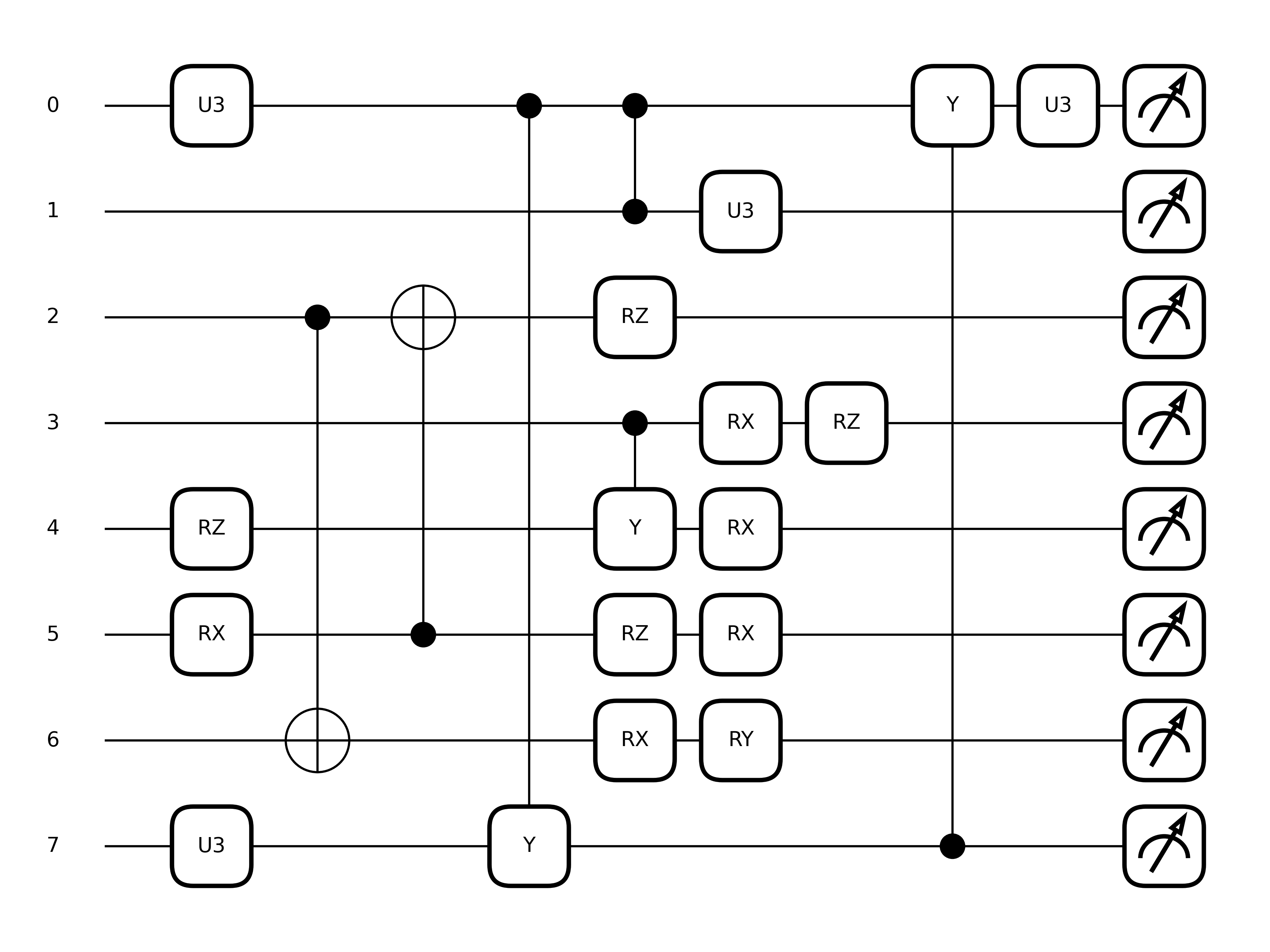}
    \caption{8-qubit state preparation}
    \end{subfigure}
    \hfill
    \begin{subfigure}[b]{0.275\textwidth}
    \centering
    \includegraphics[width=\textwidth,  height=0.75\textwidth]{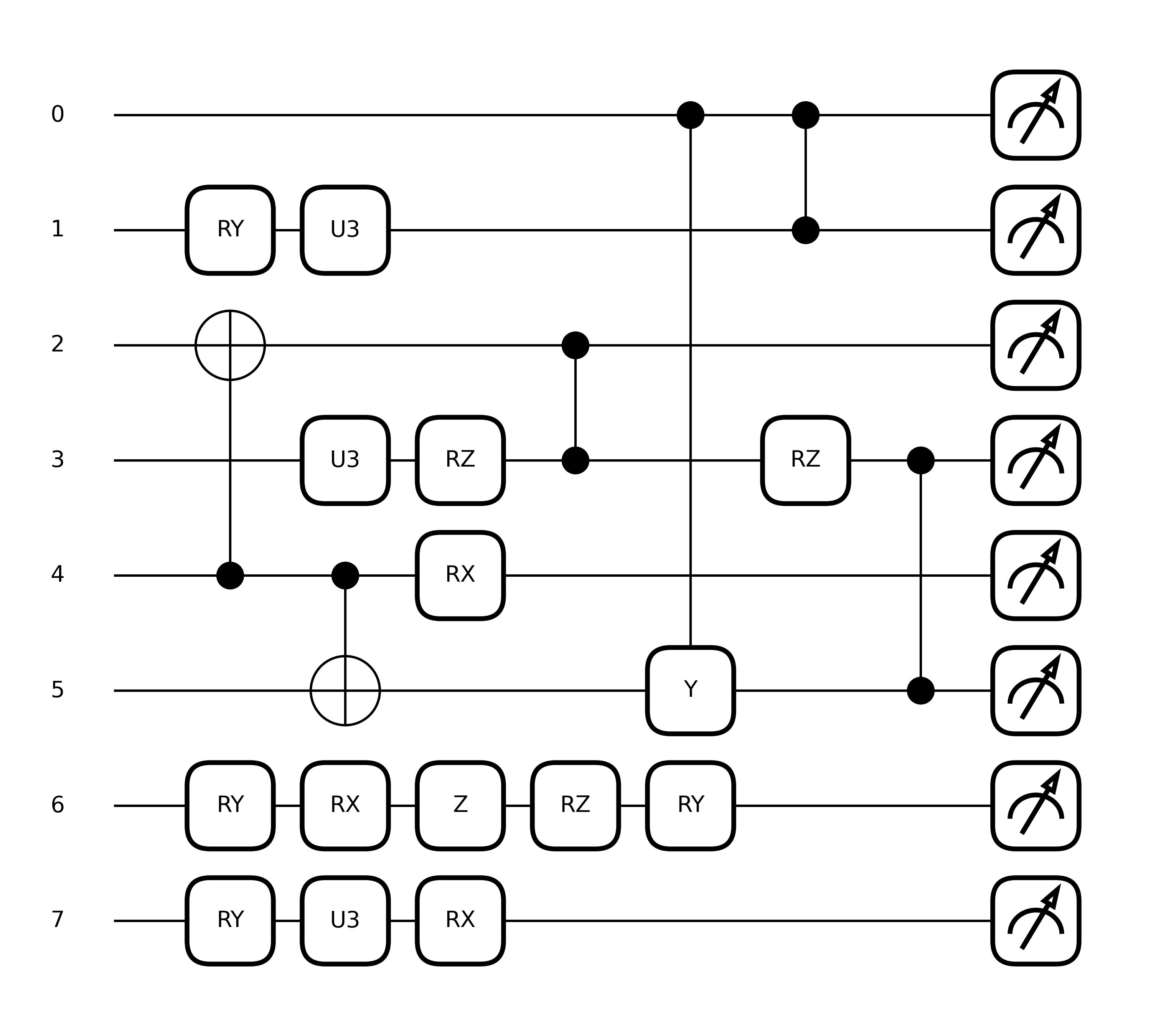}
    \caption{8-qubit max-cut}
    \end{subfigure}
    \hfill
    \begin{subfigure}[b]{0.275\textwidth}
    \centering
    \includegraphics[width=\textwidth, height=0.75\textwidth]{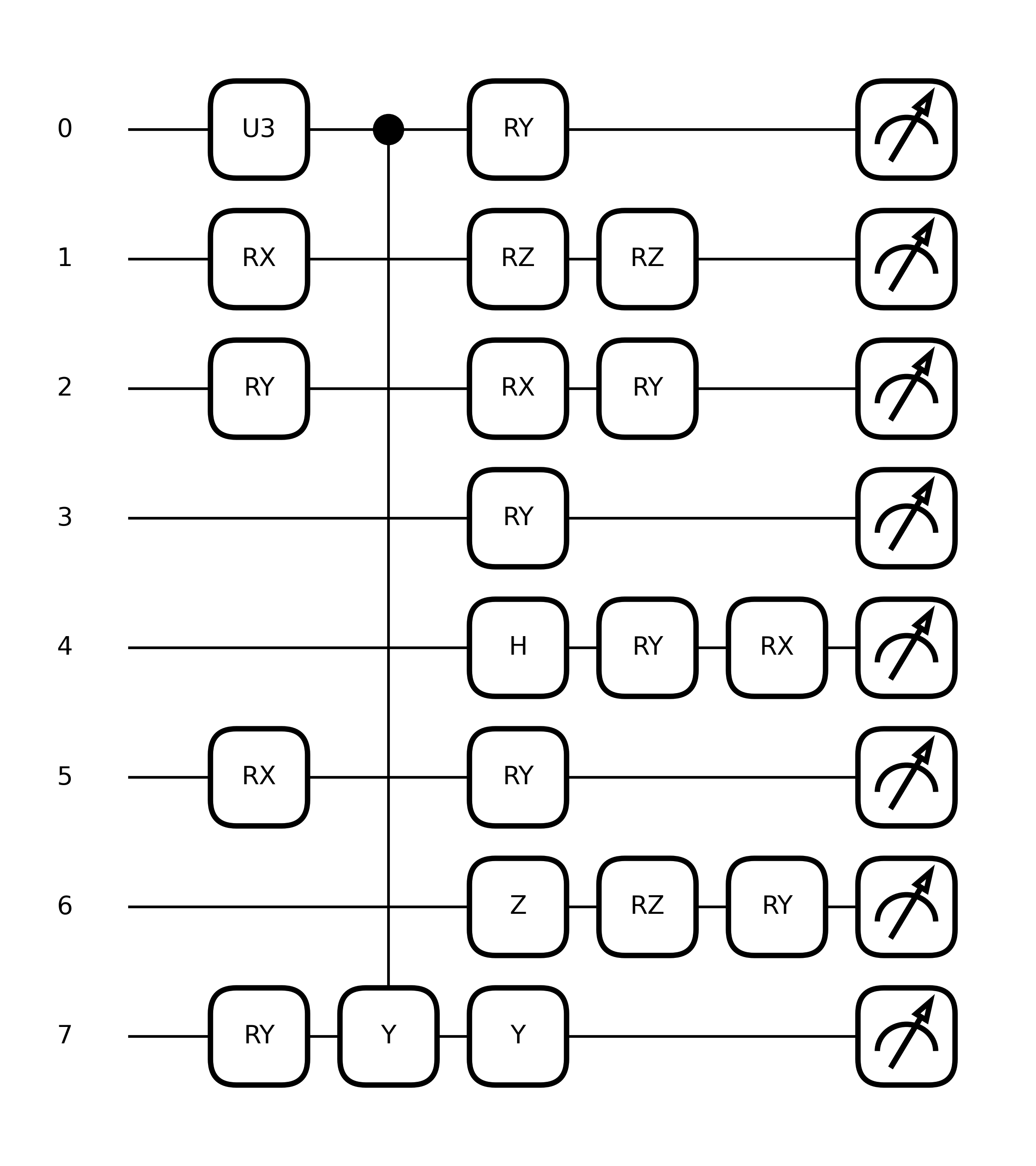}
    \caption{8-qubit quantum chemistry}
    \end{subfigure}
\end{center}
\caption{Best candidates of the six experiments in 50 runs.}
\label{best_candidates}
\end{figure*}

\clearpage
\FloatBarrier

\end{document}